\documentclass[graybox]{svmult}

\usepackage[utf8]{inputenc}
\usepackage{hyperref}
\hypersetup{
    colorlinks=true,       
    linkcolor=black,          
    citecolor=blue,        
    filecolor=blue,      
    urlcolor=blue          
}

\usepackage{mathptmx}       
\usepackage{helvet}         
\usepackage{courier}        
\usepackage{type1cm}        
\usepackage{makeidx}         
\usepackage{graphicx}        
\usepackage{multicol}        
\usepackage[bottom]{footmisc}

\usepackage{bm,subfig,amsmath,amssymb,amsfonts,mhchem}

\usepackage[usenames,dvipsnames]{xcolor}

\usepackage{multirow}

\usepackage{epstopdf}

\makeindex             

\begin{document}

\title*{Membrane-mediated interactions}
\author{Anne-Florence Bitbol, Doru Constantin and Jean-Baptiste Fournier}
\institute{Anne-Florence Bitbol \at Sorbonne Université, CNRS, Laboratoire Jean Perrin (UMR 8237), Paris, France (previous address: Lewis-Sigler Institute for Integrative Genomics and Department of Physics, Princeton University, Princeton, NJ, USA), \email{anne-florence.bitbol@sorbonne-universite.fr}
\and Doru Constantin \at Laboratoire de Physique des Solides, CNRS, Univ. Paris-Sud, Université Paris-Saclay, Orsay, France, \email{doru.constantin@u-psud.fr}
\and Jean-Baptiste Fournier \at Laboratoire “Matière et Systèmes Complexes” (MSC), UMR 7057 CNRS, Université Paris 7 Diderot, Paris Cedex 13, France, \email{jean-baptiste.fournier@univ-paris-diderot.fr}}
%
%
\setcounter{secnumdepth}{3}
\setcounter{tocdepth}{3} 


\maketitle

\abstract{Interactions mediated by the cell membrane between inclusions, such as membrane proteins or antimicrobial peptides, play important roles in their biological activity. They also constitute a fascinating challenge for physicists, since they test the boundaries of our understanding of self-assembled lipid membranes, which are remarkable examples of two-dimensional complex fluids. Inclusions can couple to various degrees of freedom of the membrane, resulting in different types of interactions. In this chapter, we review the membrane-mediated interactions that arise from direct constraints imposed by inclusions on the shape of the membrane. These effects are generic and do not depend on specific chemical interactions. Hence, they can be studied using coarse-grained soft matter descriptions. We deal with long-range membrane-mediated interactions due to the constraints imposed by inclusions on membrane curvature and on its fluctuations. We also discuss the shorter-range interactions that arise from the constraints on membrane thickness imposed by inclusions presenting a hydrophobic mismatch with the membrane.}

\section{Introduction}
\label{Intro}

Although membrane proteins were traditionally described as free to diffuse in the cell membrane~\cite{Singer72}, it was soon acknowledged that the lipid bilayer can influence their organization and thus have an impact on many aspects of their activity~\cite{Sackmann84}. Hence, interactions between proteins and the host membrane, as well as the resulting protein-protein interactions, have become fundamental topics in biophysics.

Membrane inclusions such as proteins can couple to various degrees of freedom of the membrane (curvature, thickness, composition, tilt, etc.), thus giving rise to several types of membrane-mediated interactions. It is noteworthy that these interactions are often non-specific, i.e. they do not involve the formation of chemical bonds between the various components. Thus,  understanding these interactions calls for a description of the membrane as a self-assembled system whose properties are collectively determined, and not merely given by the chemical properties of the molecules involved~\cite{Jensen04}. Over the last few decades, it has become clear that the concepts developed in soft matter physics to describe self-organized systems are extremely useful in this context, and that coarse-grained effective models such as the Helfrich model of membrane elasticity~\cite{Helfrich73} can yield valuable insight.

In this chapter, we review the membrane-mediated interactions between inclusions such as membrane proteins that arise from direct  constraints imposed by these inclusions on the shape of the membrane. Our point of view is mostly theoretical, in agreement with the history of this research field, but we also discuss the numerical and experimental results that are available. For clarity, we treat separately the effects that result from the coupling of the inclusions with membrane curvature and those that arise from their coupling with membrane thickness. Note however that a given inclusion can couple to both of these degrees of freedom. The first case, presented in Section~\ref{SecAvDef}, leads to interactions with a much larger range than the characteristic size of the inclusions, which will be referred to as ``long-range interactions''. Such effects can be described starting from the coarse-grained Helfrich model~\cite{Helfrich73}. The second case, discussed in Section~\ref{sec:short}, yields a much shorter-range interaction, and requires more detailed effective models of the membrane.

Other types of membrane-mediated interactions, arising from other underlying membrane degrees of freedom such as lipid composition and tilt, will not be discussed in detail. Besides,  important applications such as the crystallization of membrane proteins and the interaction between constituents of such crystals, are outside of the scope of this chapter.

\section{Long-range membrane-mediated interactions}
\label{SecAvDef}

Inclusions such as proteins are generally more rigid than the membrane. Therefore, they effectively impose constraints on the shape of the membrane, especially on its curvature, which plays a crucial part in membrane elasticity. These constraints in turn yield long-range membrane-mediated interactions between inclusions.

We will review the first theoretical predictions of these interactions, before moving on to further results in the analytically tractable regime of distant inclusions embedded in almost-flat membranes, including anisotropy, multi-body effects, and dynamics. Extensions to other geometries will then be discussed, including the compelling but tricky regime of large deformations, where numerical simulations provide useful insight. Finally, we will examine the available experimental results.

\subsection{First predictions}

\subsubsection{Seminal paper}

The existence of long-range membrane-mediated forces between inclusions in lipid membranes was first predicted in Ref.~\cite{Goulian93etc}. The curvature elasticity of the membrane was described by the tensionless Helfrich Hamiltonian~\cite{Helfrich73}. For an up-down symmetric membrane, it reads:
\begin{equation}
H=\int dA\left[\frac{\kappa}{2}\left(c_1+c_2\right)^2+\bar\kappa\,c_1 c_2\right]\,,
\label{HelfCov}
\end{equation}
where $\kappa$ is the bending rigidity of the membrane and $\bar\kappa$ is its Gaussian bending rigidity, while $c_1$ and $c_2$ denote the local principal curvatures of the membrane, and $A$ its area. This elastic energy penalizes curvature. For small deformations of the membrane around a planar shape, Eq.~\ref{HelfCov} can be approximated by
\begin{equation}
H[h]=\int d\bm{r}\left\{\frac{\kappa}{2}\left[\nabla^2 h(\bm{r})\right]^2+\bar\kappa\,\det[\partial_i\partial_j h(\bm{r})]\right\}\,,
\label{HelfSansSigma}
\end{equation}
where $h(\bm{r})$ is the height of the membrane at position $\bm{r}=(x,y)\in\mathbb{R}^2$ with respect to a reference plane, and $(i,\,j)\in\{x,y\}^2$. The Hamiltonian in Eq.~\ref{HelfSansSigma} is massless and features a translation symmetry ($h\rightarrow h+C$ where $C$ is independent of position) that is broken in a ground-state configuration, yielding Goldstone modes. The associated long-range correlations give rise to long-range membrane-mediated interactions. Neglecting the effect of the membrane tension $\sigma$, as in Eqs.~\ref{HelfCov} and~\ref{HelfSansSigma}, is legitimate below the length scale $\sqrt{\kappa/\sigma}$. Note that the simplified Hamiltonian in Eq.~\ref{HelfSansSigma} is quadratic in the field $h$, i.e. the field theory is Gaussian. 

\begin{figure}[htb] 
\centering
\includegraphics[width=0.5\textwidth]{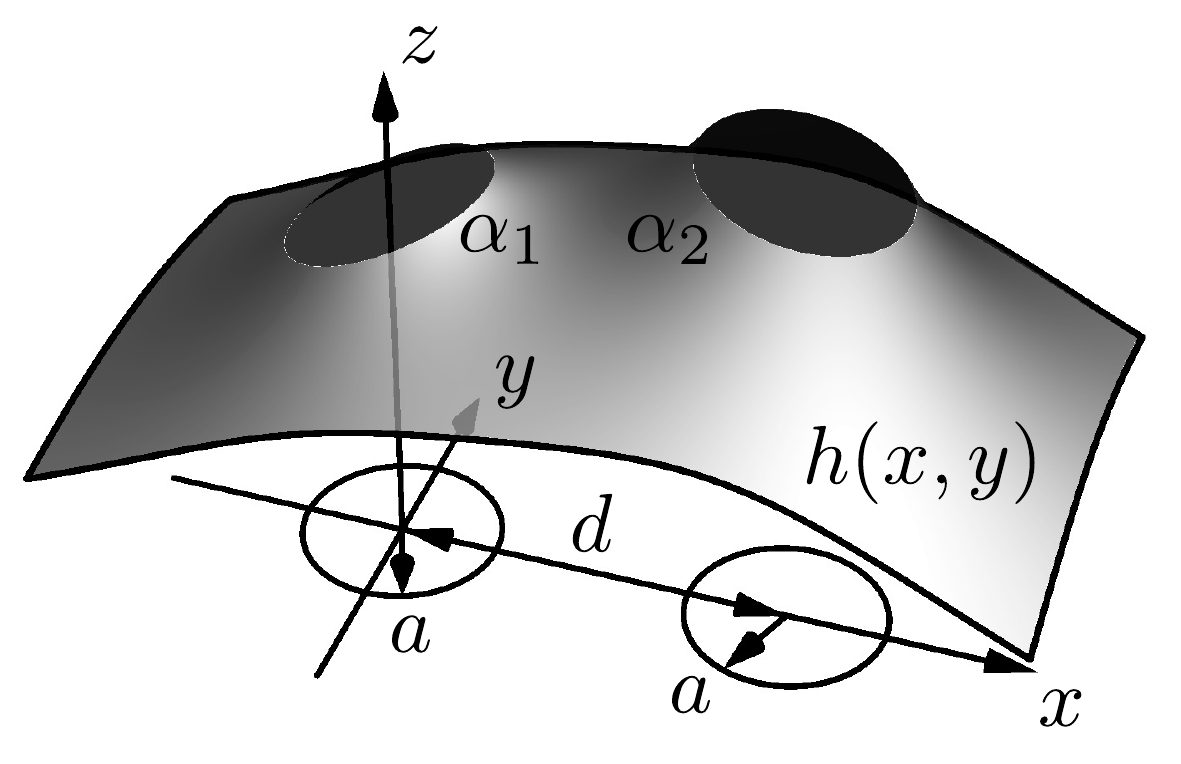}
\caption{Ground-state shape of a membrane containing two rigid disk-shaped inclusions that impose the contact angles $\alpha_1$ and $\alpha_2$, obtained by solving the Euler-Lagrange equation (see Ref.~\cite{Fournier15}). The membrane shape is described by its height $h$ with respect to the plane $z=0$. The radius of the inclusions is denoted by $a$, and the center-to-center distance by $d$.\label{Gou}}
\end{figure}

In Ref.~\cite{Goulian93etc}, inclusions are characterized by bending rigidities different from those of the membrane bulk. A zone with slightly different rigidities can represent a phase-separated lipid domain, while a very rigid zone can represent a protein. Both regimes (perturbative and strong-coupling) are discussed, in the geometry of two identical circular domains of radius $a$ at large separation $d\gg a$ (see Fig.~\ref{Gou}). An interaction potential proportional to $1/d^4$ is obtained in both regimes. 

In the perturbative regime, the interaction depends on the perturbations of $\kappa$ and $\bar\kappa$ in the inclusions and on the value of $\kappa$ in the membrane, as well as on $k_\mathrm{B}T$.

Besides, a low-temperature interaction is obtained for rigid inclusions that impose a contact angle with the membrane, e.g. cone-shaped inclusions~\cite{Goulian93etc, Fournier97comment}:
\begin{equation}
U_1(d)=4\pi\kappa(\alpha_1^2+\alpha_2^2)\,\frac{a^4}{d^4}\,,
\label{ElastGou}
\end{equation}
where $\alpha_1$ and $\alpha_2$ are the contact angles imposed by inclusion 1 and inclusion 2 (see Fig.~\ref{Gou}). This interaction is obtained by calculating the membrane shape that minimizes the membrane curvature energy in Eq.~\ref{HelfSansSigma} in the presence of the inclusions. It arises from the ground-state membrane deformation due to the inclusions, and vanishes for up-down symmetric inclusions. It is repulsive. Note that this interaction does not depend on the Gaussian bending rigidity of the membrane~\cite{Fournier97comment}, in contrast with the perturbative case~\cite{Goulian93etc}. Indeed, the Gaussian curvature energy term only depends on the topology of the membrane and on boundary conditions~\cite{Fournier97comment}. Hence, in most subsequent studies of the membrane-mediated forces between rigid membrane inclusions, the Gaussian curvature term in Eq.~\ref{HelfSansSigma} is discarded.

Another interaction, which is attractive and originates from the thermal fluctuations of the membrane shape, was predicted as well between rigid inclusions~\cite{Goulian93etc, Golestanian96b}:
\begin{equation}
U_2(d)=-6\,k_\mathrm{B}T\,\frac{a^4}{d^4}\,.
\label{CasiGou}
\end{equation}
Importantly, this fluctuation-induced interaction is independent of elastic constants and of contact angles. It exists even for up-down symmetric inclusions (imposing $\alpha_1=0$ and $\alpha_2=0$) that do not deform the ground-state membrane shape.

Multipole expansions valid for $a\ll d$ were used to calculate these interactions for rigid inclusions. Details on these expansions are presented in Refs.~\cite{Golestanian96b,Fournier15}. Only the leading-order terms in $a/d$ were obtained in Ref.~\cite{Goulian93etc}. This method was recently pushed further, yielding higher-order terms in $a/d$~\cite{Fournier15}.

\subsubsection{Point-like approach}

Ref.~\cite{Park96} extended the study of of Ref.~\cite{Goulian93etc}. Membrane elasticity was described by Eq.~\ref{HelfSansSigma} as in Ref.~\cite{Goulian93etc}, but different membrane-inclusion couplings were considered. Rigid inclusions were treated through a coupling Hamiltonian favoring a relative orientation of their main axis and of the normal of the membrane. The membrane-mediated interaction was calculated in the limit of very small inclusions, where the ultraviolet cutoff of the theory $\Lambda$ appears. The radius $a$ of the inclusions was related to $\Lambda$ through $\Lambda=2/a$~\cite{Park96}, yielding agreement with the results of~\cite{Goulian93etc}: the total interaction energy obtained is the sum of $U_1$ and $U_2$ (Eqs.~\ref{ElastGou} and~\ref{CasiGou}). 

This opened the way to direct point-like descriptions of membrane inclusions. In Ref.~\cite{Netz97}, a perturbative approach was taken, where the coupling with the membrane and the inclusions was assumed to be linear or quadratic in the local mean curvature at the point location of the inclusion. In Ref.~\cite{Kim98}, the insertion energy of a protein in the membrane was approximated by a term proportional to the Gaussian curvature of the membrane at the insertion point. Then, in Refs.~\cite{Dommersnes99, Dommersnes99b}, inclusions were modeled as more general local constraints on the membrane curvature tensor. Considering inclusions as point-like is justified in the case of membrane proteins, since their typical radius is comparable to membrane thickness, which is neglected when the membrane is considered as a surface, as in Eq.~\ref{HelfSansSigma}. This description simplifies the calculation of membrane-mediated interactions, by eliminating the need for a multipole expansion. In practice, one writes the partition function of the membrane described by the elastic energy in Eq.~\ref{HelfSansSigma} (discarding Gaussian curvature), modeling inclusions as point curvature constraints~\cite{Dommersnes99, Dommersnes99b}. For one inclusion imposing a local isotropic curvature $c$ in $\bm{r}_0$, these constraints read $\partial ^2_x h(\bm{r}_0)=\partial ^2_y h(\bm{r}_0)=c$ and $\partial _x \partial _y h(\bm{r}_0)=0$. Then, the part of the free energy that depends on the distance $d$ between the inclusions is the sum of $U_1$ and $U_2$ (Eqs.~\ref{ElastGou} and~\ref{CasiGou}), where the effective radius $a$ of the point-like inclusions appears through the cutoff $\Lambda=2/a$, and the effective contact angle is $\alpha=a c$. 

Refs.~\cite{Yolcu11,Yolcu12} formalized the connection between the original description of inclusions as rigid objects~\cite{Goulian93etc}, and the more convenient point-like description.  The effective field theory formalism developed in Refs.~\cite{Yolcu11,Yolcu12} for membranes (see also Ref.~\cite{Yolcu12b} for fluid interfaces, and Ref.~\cite{Yolcu14} for a review) considers inclusions as point-like particles, and captures their structure and the boundary conditions they impose via localized coupling terms. In practice, a series of generic scalar localized terms consistent with the symmetries of the system is added to the curvature energy describing the bare membrane. Each term in the series is polynomial in the derivatives of the membrane height $h$, taken at the point position of the inclusion. The coefficients of each term of the series are then obtained by matching observables, such as the ground-state membrane shape responding to an imposed background, between the full model with extended inclusions and the effective field theory~\cite{Yolcu12}. These Wilson coefficients are analogous to charges, polarizabilities etc. of the inclusions and describe the interplay between the membrane and the inclusions, by encoding the long-range effects of short-range coupling~\cite{Yolcu14}. Membrane-mediated interactions can be obtained from this effective field theory. It gives back the leading terms in $a/d$ obtained previously, with a generalization to inclusions with different radii, and yields higher-order corrections~\cite{Yolcu11,Yolcu12}. This general and powerful method could be extended to complex inclusions with specific Wilson coefficients, and also enables general derivation of scaling laws through power counting. However, one should bear in mind that its existing application to rigid disk-shaped inclusions \textit{a priori} yields results specific to this particular model of the inclusions. In particular, the discrepancy obtained with previous point-like approaches on certain higher-order terms~\cite{Yolcu12} should be regarded as a different result obtained for a different model, since previous point-like approaches did not aim to fully mimic rigid disk-shaped inclusions. Note that higher-order terms were recently calculated in the framework of extended disks~\cite{Fournier15}, showing agreement with~\cite{Yolcu12} and pushing the expansion further. 

\subsubsection{Two types of interactions}

The long-range membrane-mediated interaction between rigid inclusions comprises two leading-order terms that both depend on the fourth power of $a/d$ (Eqs.~\ref{ElastGou} and~\ref{CasiGou})~\cite{Goulian93etc}. Subsequent works~\cite{Park96,Dommersnes99,Dommersnes99b,Yolcu11,Yolcu12} demonstrated that the total interaction is the sum of these two terms, one coming from the ground-state deformation of the membrane by the inclusions (Eq.~\ref{ElastGou}) and the second one arising from entropic effects (Eq.~\ref{CasiGou}). However, it should be noted that the separation of these two terms is mostly of formal interest, since the ground-state shape, which is obtained by minimizing the Hamiltonian of the system, may not be of much practical relevance. In practice, one may be able to measure experimentally the average shape of a membrane, but in general it would not coincide with the ground-state one, except in the regime of small deformations. In this regime, which has been the focus of most theoretical work, the membrane Hamiltonian is quadratic (Eq.~\ref{HelfSansSigma}): then, the separation of the two terms makes sense. Let us now discuss each of these two terms.

The first term, $U_1$ (Eq.~\ref{ElastGou}), arises from the interplay of the ground-state deformations of the membrane due to the presence of each of the inclusions, and it was first obtained in Ref.~\cite{Goulian93etc} by taking the (fictitious) zero-temperature limit. It also corresponds to the membrane-mediated interaction within a mean-field approximation.

The second term, $U_2$ (Eq.~\ref{CasiGou}), is a fluctuation-induced or entropic effect, which exists even if both inclusions impose vanishing contact angles. Remarkably, in the case of rigid inclusions, the only energy scale involved is $k_\mathrm{B}T$: this interaction is universal. It arises from the constraints imposed by the inclusions on the thermal fluctuations of the shape of the membrane, which is a field with long-range correlations. It is analogous to the Casimir force in quantum electrodynamics (see e.g.~\cite{Park96,Golestanian96c,Golestanian96b,Dommersnes99,Dommersnes99b}), which arises from the constraints imposed by non-charged objects (e.g. metal plates) on the quantum fluctuations of the electromagnetic field~\cite{Casimir48,Milonni}. This fluctuation-induced interaction is thus often termed ``Casimir'' or ``Casimir-like''. In Ref.~\cite{Helfrich01}, the fluctuation-induced force between membrane inclusions was recovered from the entropy loss associated to the suppression of fluctuation modes, thus reinforcing the formal analogy with the Casimir force. Fluctuation-induced forces analogous to the Casimir force exist in several other soft matter systems, where thermal fluctuations play an important part~\cite{Kardar99,Gambassi09}. They were first discussed by Fisher and de Gennes in the context of critical binary mixtures~\cite{Fisher78}. This ``critical Casimir'' force has been measured experimentally between a colloid and a surface immersed in a critical binary mixture~\cite{Hertlein08}. Interestingly, such critical Casimir forces have been predicted to exist in membranes close to a critical point in lipid composition, and that they are very long-range, with power laws up to $(a/d)^{1/4}$~\cite{Machta12}. Their sign depends on the boundary conditions imposed by the inclusions~\cite{Machta12}, as in the three-dimensional critical case~\cite{Gambassi09}.

Let us now compare the magnitude of these two types of interactions. For two identical inclusions imposing the same contact angle $\alpha$, the interactions in Eq.~\ref{ElastGou} and~\ref{CasiGou} have the same modulus if 
\begin{equation}
|\alpha|=\sqrt{\frac{3}{4\,\pi}\,\frac{k_\mathrm{B}T}{\kappa}}\,.
\end{equation}
Using the typical value $\kappa\approx25\,k_\mathrm{B}T$ gives $|\alpha|\approx 6^\circ$: for larger contact angles, the mean-field repulsion dominates over the fluctuation-induced attraction.

\subsection{Further developments on distant inclusions embedded in almost-flat membranes}

\subsubsection{In-plane anisotropy}

\begin{figure}[htb] 
\centering
\includegraphics[width=\textwidth]{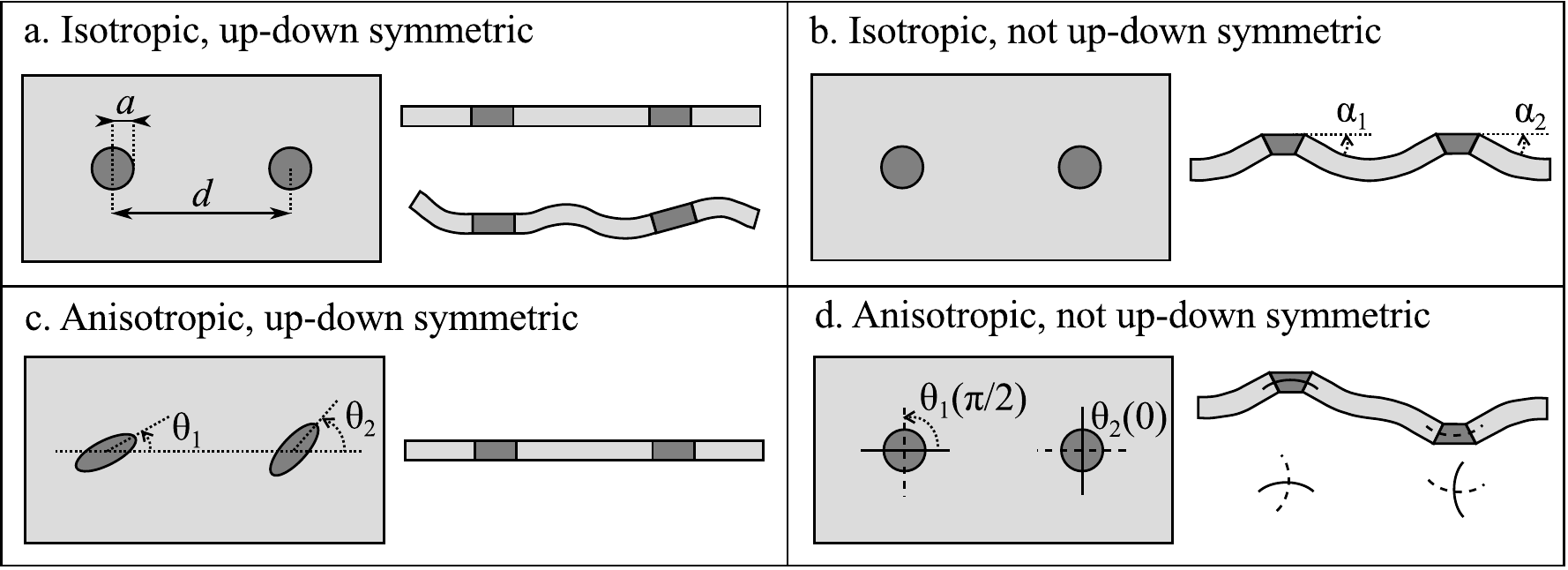}
\caption{Schematic representation of the different cases for inclusions with separation $d$ much larger than their characteristic size $a$, embedded in a membrane with small deformations around the flat shape. In each case, a view from above and a longitudinal cut are presented. Thermal fluctuations of the shape of the membrane are only represented in the bottom right cut of panel a.\label{Cases}}
\end{figure}

Until now, we discussed the simple case of two inclusions with isotropic (i.e. disk-shaped) in-plane cross-section, which was the first case investigated~\cite{Goulian93etc}.  However, real membrane inclusions, such as proteins, have various shapes. Fig.~\ref{Cases} shows a schematic of the different cases at stake: those in panels a and b were discussed above, and those in panels c and d will be discussed here.

In Ref.~\cite{Park96}, the case of anisotropic cross-sections was treated through a coupling between membrane curvature and symmetric traceless tensor order parameters constructed from the main direction of the inclusion cross-section, integrated over the surface of the inclusion cross-section. The interaction energies obtained are anisotropic, and depend on $d$ as $1/d^4$ for up-down symmetric inclusions that interact only through the fluctuation-induced interaction (see Fig.~\ref{Cases}c), just as in the case of isotropic cross-sections. However, inclusions that break the up-down symmetry of the membrane feature an anisotropic interaction with a stronger $1/d^2$ power law. Its angle dependence is $\cos(2(\theta_1+\theta_2))$, where $\theta_i$ is the angle between the main in-plane axis of inclusion $i$ and the line joining the two inclusion centers (Fig.~\ref{Cases}c). This orientation dependence is that of a quadrupole-quadrupole interaction~\cite{Golestanian96a,Stamou00}, and the interaction energy is minimized whenever $\theta_1+\theta_2=0$ (or equivalently $\theta_1+\theta_2=\pi$). This interaction is attractive for a wide range of relative orientations, while the analogous interaction between inclusions with an isotropic cross-section is always repulsive (see Eq.~\ref{ElastGou}). 

The in-plane anisotropic case of rigid up-down symmetric rods imposing vanishing contact angles to the membrane on their edges was treated in Refs.~\cite{Golestanian96a, Golestanian96b}. Only the fluctuation-induced interaction is then at play (as in Fig.~\ref{Cases}c). In this study, thin rods were considered in the limit of vanishing width, and in the ``distant'' regime where their length $L$ is much smaller than their separation $d$. The opposite case $d\ll L$ will be discussed in Sec.~\ref{closeRods}. The power law obtained is in $1/d^4$, as in the case of isotropic cross sections (Eq.~\ref{CasiGou}), and the only energetic scale involved in this fluctuation-induced force is $k_\mathrm{B}T$. The angular dependence of the interaction is $\cos^2[2(\theta_1+\theta_2)]$, yielding energy minima for $\theta_1+\theta_2=0$ and $\pi/2$.

Anisotropic cross-sections were revisited within the point-like approach in Refs.~\cite{Dommersnes99,Dommersnes02}. In this model, inclusions couple to the membrane by locally imposing a generic curvature tensor, with eigenvalues (principal curvatures) denoted by $K+J$ and $K-J$. The interaction between two such identical inclusions then reads, to leading order in $a/d$~\cite{Dommersnes99,Dommersnes02}:
\begin{equation}
U_3(d)=-8\pi\kappa\frac{a^4}{d^2}\left\{2J^2\cos(2(\theta_1+\theta_2))+JK\left[\cos(2\theta_1)+\cos(2\theta_2)\right]\right\}\,,
\label{anisotropic}
\end{equation}
where $\theta_i$ are angles between the line joining the inclusion centers and their axis of smallest principal curvature (see Fig.~\ref{Cases}d). This term vanishes for isotropic inclusions ($J=0$), consistently with Refs.~\cite{Goulian93etc,Park96}. Furthermore, in the fully anisotropic case $K=0$, corresponding to a saddle, the power law and the angular dependence both agree with the up-down symmetry breaking and anisotropic cross-section case of Ref.~\cite{Park96}. Eq.~\ref{anisotropic} shows that in the generic case where $J$ and $K$ are nonzero, the angular degeneracy of the lowest-energy state is lifted, and (assuming without loss of generality that $K$ and $J$ have the same sign) the inclusions tend to align their axis of smallest principal curvature along the line joining their centers. Their interaction is then attractive~\cite{Dommersnes99}. This interaction (Eq.~\ref{anisotropic}) was recovered in Ref.~\cite{Yolcu11} (with different angle notations), and generalized to inclusions with different radii. 

Subleading terms in $1/d^4$ were also calculated in Refs.~\cite{Dommersnes99} and~\cite{Yolcu11}, featuring different results (as for the subleading terms in the isotropic case). One should keep in mind that the models at stake are different, since Ref.~\cite{Dommersnes99} considers fully point-like inclusions while Ref.~\cite{Yolcu11} models disk-shaped ones with finite radius through the effective field theory. While the agreement of these models on the leading-order term is a nice sign of robustness, there is no reason to expect an exact agreement at all orders.

Ref.~\cite{Dommersnes99} also investigated the fluctuation-induced interaction, but its leading-order term was found not to be modified with respect to the isotropic case (Eq.~\ref{CasiGou}). This is at variance with the anisotropy obtained in Refs.~\cite{Golestanian96a, Golestanian96b} for the flat rods, but one should keep in mind that the point-like saddles do not correspond to the limit of the distant flat rods. 

\subsubsection{Multi-body effects and aggregation}

A crucial and biologically relevant question is how long-range membrane-mediated interactions drive the collective behavior of inclusions, in particular aggregation. One would be tempted to start by summing the pairwise potentials discussed above, but these long-range membrane-mediated interactions are not pairwise additive. Non-pairwise additivity is a general feature of fluctuation-induced interactions. For instance, the existence of a three-body effect in the van der Waals--London interaction was demonstrated in Ref.~\cite{Axilrod}. The interaction due to the ground-state membrane deformation is not additive either. Indeed, if one considers inclusions that impose boundary conditions to the membrane on their edges, a shape minimizing the energy in the presence of one inclusions will generically not satisfy the boundary conditions imposed by the other one, yielding non-additivity~\cite{Yolcu14}.

Three-body and four-body long-range membrane-mediated interactions were first calculated within a perturbative height-displacement model, breaking up-down symmetry but retaining in-plane anisotropy, in Ref.~\cite{Park96}. The distance dependence of the three-body term involves terms in $1/(d_{12}^2d_{23}^2)$ where $d_{ij}$ is the distance between particles $i$ and $j$. These interactions were also investigated in Ref.~\cite{Netz97}, in a different perturbative approach, considering in particular inclusions that favor a given average curvature, and then in Ref.~\cite{Kim98} in a point-like framework, but this particular calculation was recently shown to miss some contributions~\cite{Yolcu11}. 

\begin{figure}[htb] 
\centering
\includegraphics[width=0.5\textwidth]{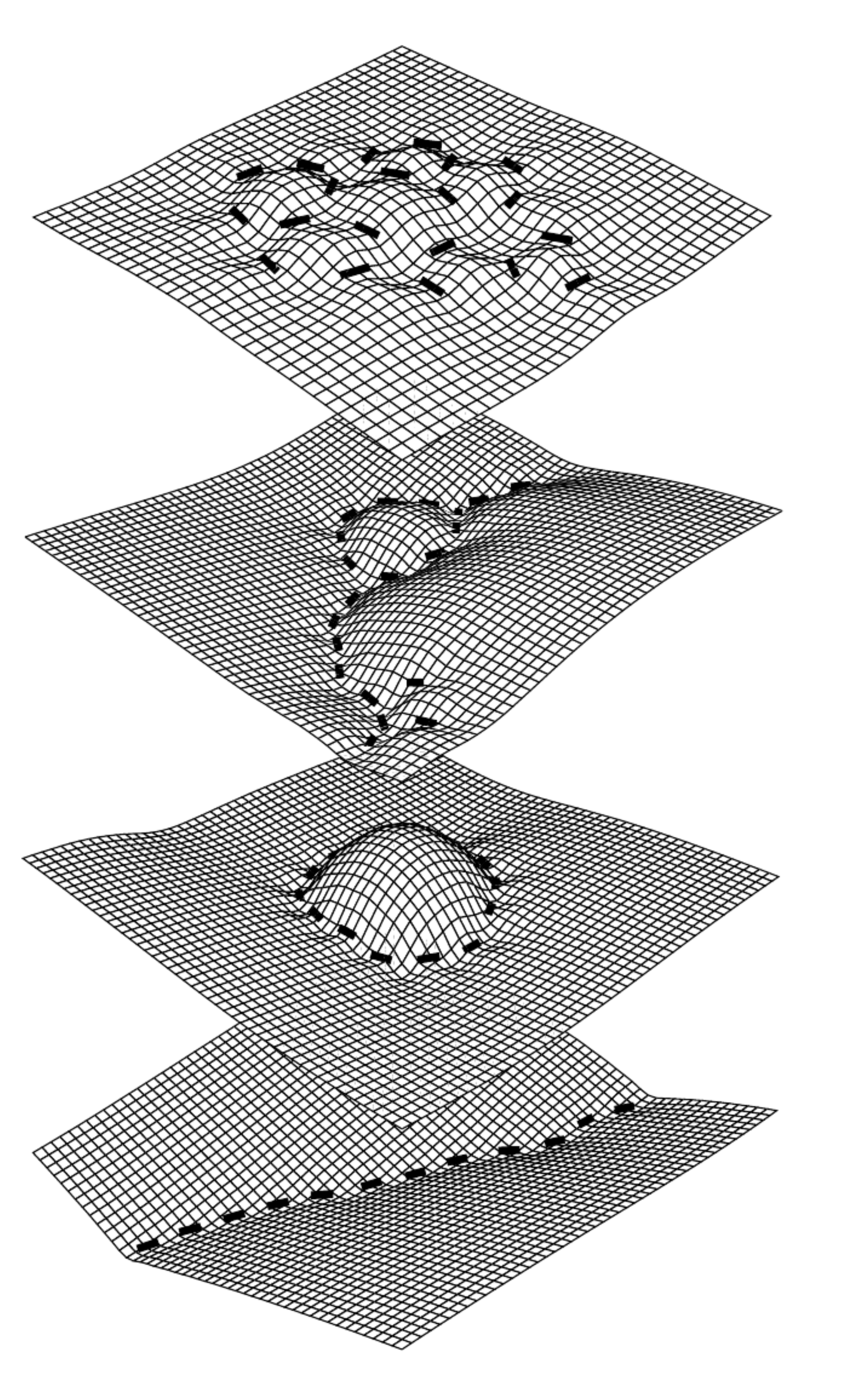}
\caption{Typical equilibrium aggregates obtained from Monte-Carlo simulation of 20 identical point-like anisotropic curvature-inducing inclusions. Each panel represents a different set of $(J,K)$ values. Reproduced from Ref.~\cite{Dommersnes99}. \label{Aggr}}
\end{figure}

In Ref.~\cite{Dommersnes99}, the multi-body interactions and the aggregation of point-like inclusions locally imposing a curvature tensor were investigated. This generic model can include both up-down symmetry breaking and in-plane anisotropy depending on the curvature tensor imposed. The leading three-body interaction was found to involve terms in $1/(d_{12}^2d_{23}^2)$, as in Ref.~\cite{Park96}, and to vanish for inclusions imposing a zero curvature tensor~\cite{Dommersnes99}. Monte-Carlo simulations including the full multi-body interactions were performed, allowing to study the phase diagram of the system (see Fig.~\ref{Aggr}). Polymer-like linear aggregates were obtained for sufficient values of $K$ and $J$, as predicted from the leading pairwise term (Eq.~\ref{anisotropic}). A gas phase was found for small $J$, consistent with the fact that for isotropic inclusions ($J=0$) that break the up-down symmetry ($K\neq 0$), the leading pairwise interaction is repulsive (Eq.~\ref{ElastGou}). Finally, for small $K$ and large $J$, aggregates were obtained, some of which had an ``egg-carton'' structure. This is made possible by the angular degeneracy of the lowest-energy state for $K=0$ in the leading pairwise term (Eq.~\ref{anisotropic}). Multi-body interactions were shown to be quantitatively important, but the effect of the fluctuation-induced interaction (Eq.~\ref{CasiGou}) was found to be negligible~\cite{Dommersnes99}. The analytical calculation of multi-body effects was performed in this framework in Ref.~\cite{Dommersnes02}, where the ``egg-carton'' aggregates were also further studied and related to experimentally-observed structures. 

Coarse-grained molecular-dynamics simulations of the highly anisotropic curvature-inducing N-BAR domain proteins adhering on membranes have demonstrated linear aggregation of these proteins on the membrane. This is a first self-assembly step, which then yields the formation of meshes enabling budding~\cite{Simunovic13}. This is qualitatively in good agreement with the predictions of Ref.~\cite{Dommersnes99}.

The influence of the long-range elastic repulsion between isotropic inclusions that break the up-down symmetry of the membrane on their aggregation was also discussed in Ref.~\cite{Destainville08}, but within a less specific framework including other types of interactions. In this work, this repulsive interaction (Eq.~\ref{ElastGou}) plays the role of an energetic barrier to aggregation.

In Ref.~\cite{Weikl01}, the collective behavior of inclusions locally penalizing local curvature (either only mean curvature or also Gaussian curvature) was studied using a mean-field theory for the inclusion concentration and Monte-Carlo simulations. Since the inclusions considered retain both up-down symmetry and in-plane isotropy, the only membrane-mediated interaction at play is an attractive fluctuation-induced one similar to that in Eq.~\ref{CasiGou}. Direct interactions were also included. Aggregation was found to occur even for vanishing direct interactions, provided that the rigidity of the inclusions was sufficient~\cite{Weikl01}. Hence, fluctuation-induced interactions may be relevant for aggregation, at least in the absence of other, stronger, interactions. Note that Eq.~(\ref{CasiGou}) shows that the amplitude of fluctuation-induced interactions is quite small. For instance, $d=4a$ yields $U_2\approx 0.02\, k_\mathrm{B}T$ (all the results discussed so far are strictly relevant only for $d\gg a$).

 In Ref.~\cite{Yolcu11}, the general effective field theory framework was used in the case of in-plane isotropic inclusions. The leading-order and next-order three-body interaction terms due to the ground-state membrane deformation between up-down symmetry-breaking inclusions were obtained, as well as the leading three-body and four-body fluctuation-induced interactions. 

\subsubsection{Membrane tension}

Until now, we have focused on the regime where bending rigidity dominates over membrane tension. This is appropriate for length scales below $\sqrt{\kappa/\sigma}$. As $\sigma$ is in the range $10^{-6}-10^{-8}$~N/m for floppy membranes, while $\kappa\simeq10^{-19}$~J, this length scale is then of order 1~$\mu$m. However, membrane tensions can span several orders of magnitude~\cite{Rawicz00} depending on external conditions (e.g. osmotic pressure), so it is relevant to go beyond $\sqrt{\kappa/\sigma}$. For small deformations around a planar shape, the quadratic Hamiltonian of a membrane including tension reads
\begin{equation}
H[h]=\int d\bm{r}\left\{\frac{\kappa}{2}\left[\nabla^2 h(\bm{r})\right]^2+\frac{\sigma}{2}\left[\bm{\nabla} h(\bm{r})\right]^2\right\}\,,
\label{HelfAvecSigma}
\end{equation}
where notations are the same as in Eq.~\ref{HelfSansSigma}, and where the Gaussian curvature term has been discarded. Note that, in a self-assembled membrane not submitted to external actions, each lipid adopts an equilibrium area. Hence, a membrane has no intrinsic surface tension (contrary to a liquid-gas interface), and stretching the membrane has an energy cost quadratic in the area variation. However, one usually considers a patch of membrane in contact with a reservoir made up by the rest of the membrane, so the tension term in Eq.~\ref{HelfAvecSigma} can be interpreted as arising from the chemical potential of this reservoir.

For length scales much larger than $\sqrt{\kappa/\sigma}$, tension dominates and Eq.~\ref{HelfAvecSigma} can be simplified into
\begin{equation}
H[h]=\frac{\sigma}{2}\int d\bm{r}\left[\bm{\nabla} h(\bm{r})\right]^2\,.
\label{HelfSltSigma}
\end{equation}
This case applies to a tense membrane at large scales, but also to a liquid interface (neglecting gravity). From a formal point of view, techniques similar to those employed in the bending-dominated case can be used, since the Hamiltonian is also quadratic with a single term. 

Let us first focus on inclusions that do not break the up-down symmetry of the membrane. In Refs.~\cite{Golestanian96a, Golestanian96b}, the fluctuation-induced interaction between two distant up-down symmetric rigid thin rods embedded in such a surface was calculated. It was found to be similar to the analogous bending-dominated case (see above), with the same $1/d^4$ power law, but with a different angular dependence.

Refs.~\cite{Lehle07,Noruzifar09} considered the tension-dominated case of ellipsoidal colloids trapped at a fluid interface. In the case where the colloid height fluctuations are included but their contact line with the fluid is pinned, long-range fluctuation-induced interactions were obtained. This case is analogous to that of rigid in-plane anisotropic membrane inclusions preserving the up-down symmetry. Interestingly, the power law obtained was found to depend on whether or not in-plane orientational fluctuations of the colloids were allowed. If they are not allowed, the result of Refs.~\cite{Golestanian96a, Golestanian96b} with the $1/d^4$ power law is recovered in the limit of full anisotropy. If they are allowed, a weaker anisotropic interaction with $1/d^8$ power law is obtained~\cite{Noruzifar09}. This strong dependence of the power law of fluctuation-induced forces on boundary conditions was confirmed in Ref.~\cite{Yolcu12b} through the effective field theory method, in the specific case of in-plane isotropic (disk-shaped) rigid inclusions~\cite{Yolcu11,Yolcu14}. In the case of membranes, the physical case should allow orientational fluctuations of the inclusions, and hence the $1/d^8$ power law should be considered. It is attractive and reads:
\begin{equation}
U_4(d)=-9k_\mathrm{B}T\frac{a^8}{d^8}\,.
\label{tensioncase}
\end{equation}
Hence, we expect a crossover between a $1/d^4$ power law (Eq.~\ref{CasiGou}) and a $1/d^8$ power law (Eq.~\ref{tensioncase}) as the tension becomes more important.

In Ref.~\cite{Lin11}, a scattering-matrix approach analogous to the one developed for the study of Casimir forces~\cite{Genet03,Emig08,Rahi09} was developed, and applied to the full Hamiltonian in Eq.~\ref{HelfAvecSigma} including both tension and bending. The focus was on disk-shaped elastic inclusions preserving the up-down symmetry, and on their fluctuation-induced interaction. The results obtained in the case of rigid inclusions were consistent with Eq.~\ref{CasiGou} in the bending-rigidity--dominated regime, and with Eq.~\ref{tensioncase} in the tension-dominated regime. Moreover, the crossover between these two regimes was studied numerically. The method developed in Ref.~\cite{Lin11} can potentially deal with more general cases, involving multiple complex inclusions. It appears to be complementary to the effective field theory method of Refs.~\cite{Yolcu11,Yolcu14}, and was more straightforward in the transition regime where both tension and bending are relevant~\cite{Lin11}.

Let us now focus on the interaction due to the ground-state deformation of the membrane. Ref.~\cite{Weikl98} studied the case of conical inclusions breaking up-down symmetry but retaining in-plane isotropy, and considered the full Hamiltonian in Eq.~\ref{HelfAvecSigma}. They showed that for non-vanishing tension, this interaction has a sign that depends on the relative orientation of the cones with respect to the membrane plane (i.e. on the signs of the angles they impose), contrary to the vanishing-tension case (see Eq.~\ref{ElastGou}). Furthermore, at long distances between inclusions, the interaction is exponentially cut off with a decay length $\sqrt{\kappa/\sigma}$ (it involves Bessel K functions). This property was confirmed in Ref.~\cite{Evans03}. Hence, at long distances, the fluctuation-induced force in Eq.~\ref{tensioncase} should dominate over the force due to the ground-state deformation. Conversely, in the case of colloids or inclusions with anisotropic cross-sections, Refs.~\cite{Stamou00} and~\cite{Yolcu12b} demonstrated the existence of a long-range interaction due to the ground-state deformation of the membrane. The leading term of this interaction is anisotropic and decays as $1/d^4$. 

In Ref.~\cite{Simunovic15}, the effect of tension on the aggregation of the highly anisotropic curvature-inducing N-BAR domain proteins adhering on membranes was investigated through coarse-grained molecular-dynamics simulations. Increasing tension was shown to weaken the tendency of these proteins to linear aggregation, in agreement with the predicted weakening of the ground-state membrane-mediated interaction.

\subsubsection{Summary of the interaction laws}

Table~\ref{Pow} presents a summary of the power laws of the leading-order term of the membrane-mediated interactions in the various situations discussed until now. 

\begin{table}[h t b]
\centering
\small
\begin{tabular}{|c|c|c|c|}
\hline
Dominant term in&&Fluctuation-induced&Interaction due to the \\
the Hamiltonian&Geometry& interaction& ground-state deformation --\\
in Eq.~\ref{HelfAvecSigma}&&& Vanishes if up-down symmetric\\
\hline
\hline
\multirow{6}{*}{Bending rigidity $\kappa$}&
Disks&$ 1/d^4$~\cite{Goulian93etc, Park96, Dommersnes99}&$ 1/d^4$~\cite{Goulian93etc, Park96, Dommersnes99}\\
&&&\\
\cline{2-4}
&Disks&$ 1/d^4$~\cite{Park96, Dommersnes99}&$ 1/d^2$~\cite{Park96, Dommersnes99}\\
&+anisotropy&&\\
\cline{2-4}
&Distant rods&$ 1/d^4$~\cite{Golestanian96a, Golestanian96b}&\\
&&&\\
\hline
\hline
\multirow{6}{*}{Tension $\sigma$}&Disks&$ 1/d^8$~\cite{Noruzifar09,Lin11}&Exponentially suppressed\\
&&&\\
\cline{2-4}
&Disks&$ 1/d^8$~\cite{Noruzifar09,Lin11}&$1/d^4$~\cite{Stamou00,Yolcu12b}\\
&+anisotropy&&\\
\cline{2-4}
&Distant rods&$ 1/d^4$~\cite{Golestanian96a, Golestanian96b}&\\
&&&\\
\hline
\end{tabular}
\normalsize
\caption[]{Summary of the power laws obtained for the leading-order terms of the two types of membrane-mediated interactions, as a function of the separation $d$ between the inclusions, in the regime of small deformations of a flat membrane and distant inclusions. Different inclusion geometries are considered. In the case labeled ``disks+anisotropy'', the anisotropy can be either in the inclusion shape (e.g. ellipsoidal~\cite{Park96}) or in the constraint it imposes (e.g. an anisotropic local curvature~\cite{Dommersnes99}).\label{Pow}}
\end{table}

\subsubsection{External forces and torques}
\label{extft}

Until now, we have only discussed cases where inclusions couple to the membrane shape through its curvature, either explicitly or implicitly (e.g. through rigidity). This is the relevant case in the absence of external forces or torques. External forces can yield local constraints directly on the height of the membrane, e.g. quadratic ones in the case of local trapping or linear ones in the case of local pulling~\cite{Netz97}. More specifically, inclusions may experience direct mechanical constraints if they are attached to the cytoskeleton, and torques in the presence of electrical fields because of their dipole moments~\cite{Dommersnes99b}. In these cases, one expects membrane-mediated interactions to be enhanced, because the ground-state deformations will generically be stronger than in the case where inclusions can freely reorient to minimize them, and because the constraints imposed on fluctuations will be stronger too. 

The case of inclusions subject to external torques was studied in Ref.~\cite{Dommersnes99b}, for point-like inclusions setting a curvature tensor, in in-plane isotropic case. Both external fields strong enough to effectively pin the orientations of the inclusions, and finite external fields that set a preferred orientation, were considered. In both cases, membrane-mediated forces are strongly enhanced, even more in the strong-field case. A logarithmic fluctuation-induced interaction was obtained, as well as an interaction due to the ground-state deformation which either scales as $1/d^2$ if the preferred orientations are the same for both inclusions, or logarithmically if they are different. Interestingly, these interactions depend on the relative orientation of the preferred curvatures set by the inclusions, while in the torque-free case, the interaction only depends on their absolute values (see Eq.~\ref{ElastGou})~\cite{Dommersnes99b}. 

In Ref.~\cite{Lehle07}, colloids at a fluid interface were considered, with different types of boundary conditions. In the case were the position of the colloids is considered to be frozen (both in height and in orientation), strong logarithmic fluctuation-induced interactions are obtained.

\subsubsection{Fluctuations of the interactions}
Until now, we have discussed the average values at thermal equilibrium of  membrane-mediated forces. Thermal fluctuations already play an important part since they are the physical origin of fluctuation-induced forces. But membrane-mediated forces themselves fluctuate as the shape of the membrane fluctuates. The fluctuations of these forces have been studied in Ref.~\cite{Bitbol10}, using the stress tensor of the membrane~\cite{Capovilla02, Fournier07}. This approach is inspired from those used previously for the fluctuations of Casimir forces~\cite{Barton91}, and of Casimir-like forces between parallel plates imposing Dirichlet boundary conditions on a thermally fluctuating scalar field~\cite{Bartolo02}. 

The case of two point-like membrane inclusions that locally impose a curvature tensor was studied in Ref.~\cite{Bitbol10}, for in-plane isotropic inclusions but including the up-down symmetry breaking case. Integrating the stress tensor on a contour surrounding one of the two inclusions allowed to calculate the force exerted on an inclusion by the rest of the system, in any shape of the membrane~\cite{Bitbol11_force}. The average of the force obtained gives back the known results Eqs.~\ref{ElastGou} and~\ref{CasiGou} that were obtained from the free energy in previous works. The variance of the force was also calculated, showing that the membrane-mediated force is dominated by its fluctuations. The distance dependence of the fluctuations, present in the sub-leading term of the variance, was also discussed. Interestingly, it shares a common physical origin with the fluctuation-induced (Casimir-like) force~\cite{Bitbol10}.

\subsection{Dynamics}

Fundamental interactions, e.g., electrostatic ones, are usually considered as instantaneous, in the sense that they propagate at a velocity much higher than that of the particles experiencing them. This is not the case for membrane-mediated interactions, as the spreading of membrane deformations involves slow dissipative phenomena. The dynamics of membrane-mediated interactions is a promising subject for future research. Studying out-of-equilibrium membrane-mediated interactions intrinsically requires taking into account the dynamics of the membrane. Taking care both of the motion of the membrane and of that of the inclusions is very difficult. Hence, the first theoretical study in this direction to our knowledge, Ref.~\cite{Fournier14}, considered two immobile inclusions  that simultaneously change conformation, i.e., that simultaneously create a source of deformation, and therefore trigger a time-dependent interaction as the membrane deformation spreads dissipatively. 

In Ref.~\cite{Fournier14}, inclusions were modeled as simple point-like sources of mean curvature that are triggered simultaneously at $t=0$. One could imagine cylindrical integral proteins such as ion channels transforming into conical ones upon receiving a chemical signal. The time-dependent Hamiltonian of these inclusions is $\mathcal{H}_\mathrm{inc}(t)=\theta(t)\sum_i B_i\nabla^2h(\bm r_i)$, with $\theta(t)$ the Heaviside step function, $B_i$ the curving strength, and $\bm r_i$ the position of inclusion~$i$. The dynamical reaction of the membrane to such a perturbation was studied in Ref.~\cite{Fournier14}.

As shown in the pioneering works of Refs.~\cite{Seifert93,Evans94}, the dominant dissipation mechanism at short length scales is the friction between the two monolayers of the membrane. The corresponding dissipated power per unit area is $b(\bm v^+-\bm v^-)^2$, where $\bm v^\pm$ are the velocities of the two lipid monolayers (the monolayers are denoted by $+$ and $-$) and $b\approx10^9~\mathrm{J\,s}/\mathrm{m}^4$ is the intermonolayer friction coefficient. In addition, the membrane is subjected to viscous forces from the bulk solvent, of viscosity $\eta\approx10^{-3}~\mathrm{J\,s}/\mathrm{m}^3$, and each monolayer behaves as a compressible fluid with elastic energy density $\frac{1}{2}k(\rho^\pm\pm e\nabla^2h)^2$. In this expression, $\rho^\pm$ are the monolayer relative excess densities (normalized by their equilibrium density), measured on the membrane mid-surface, $e\approx1~\mathrm{nm}$ is the distance between this surface and the neutral surface of the monolayers (where density and curvature effects are decoupled) and $k\approx0.1~\mathrm{J}/\mathrm{m}^2$. For most practical purposes, the two-dimensional viscosities of the monolayers can be neglected~\cite{Fournier15b}. 

Taking into account all these effects, Ref.~\cite{Fournier14} showed that the relaxation dynamics of a Fourier mode $\{h(q,t),\rho^\pm(q,t)\}$ in the membrane with two identical triggered inclusions is given, to linear order, by a set of two first-order dynamical equations:
\begin{align}
&2b\,\frac{\partial(\rho^+-\rho^-)}{\partial t}=-kq^2(\rho^+-\rho^-)+2keq^4h,
\label{eqd1}
\\
&4\eta q\,\frac{\partial h}{\partial t}=
-(\sigma q^2+\tilde\kappa q^4)h+keq^2(\rho^+-\rho^-)+F(\bm q,t),
\label{eqd2}
\end{align}
where $F(\bm q,t)$ is the Fourier transform of $-\delta\mathcal{H}_\mathrm{inc}/\delta h(\bm r,t)$, $\sigma$ is the membrane tension, and $\tilde\kappa=\kappa+2ke^2$ the bending rigidity at frozen lipid density~\cite{Seifert93}. Solving these linear differential equations for time evolution and integrating over the Fourier modes $\bm q$ yields the time-dependent membrane deformation produced by one or more inclusions. Then the force $f(t)$ exerted by one inclusion on the other can be obtained by integrating the membrane stress tensor~\cite{Capovilla02,Fournier07,Bitbol11_stress} around one inclusion.

\begin{figure}[htb] 
\centering
\includegraphics[width=\textwidth]{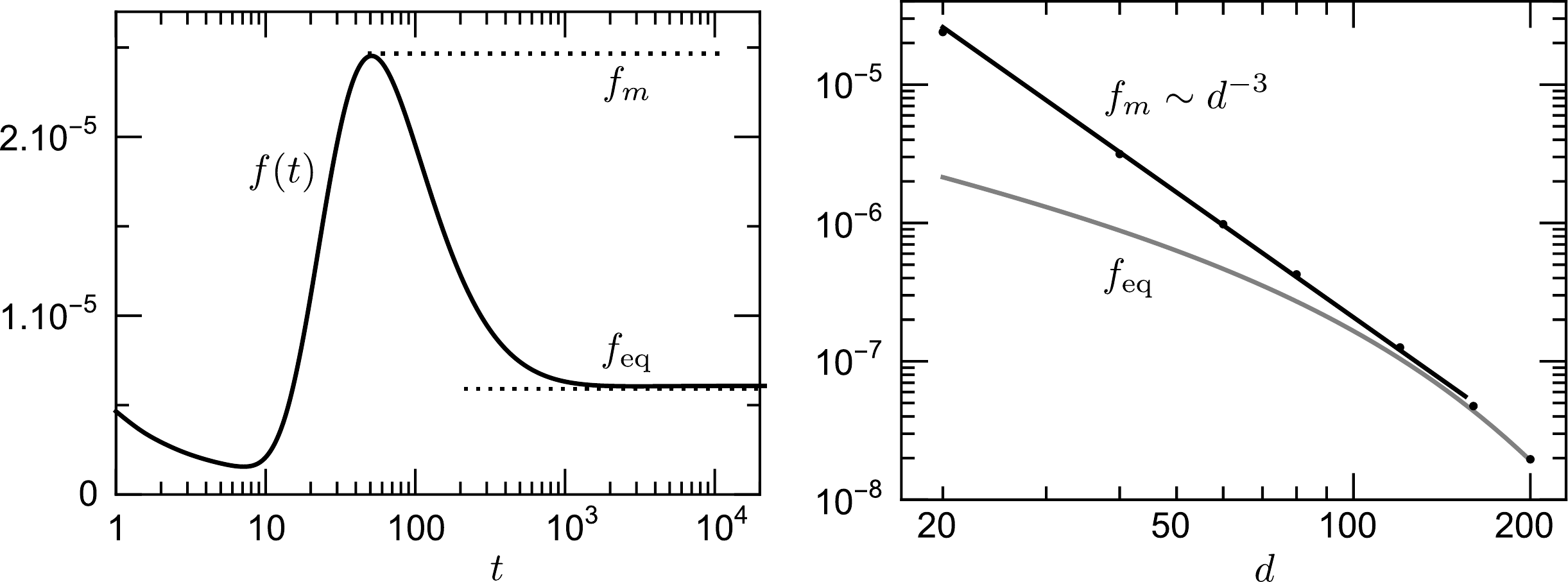}
\caption{
(a) Force $f(t)$ normalized by $B^2/(\kappa e^3)$ exchanged by two inclusions separated by $d$ versus time $t$ normalized by $4\eta e^3/\kappa\times10^3$. The parameters are $d=20 e$, $\sigma=10^{-3}\kappa/e^2$, $ke^2/\kappa=1$ and $be^2/\eta=1000$. (b) Dependence of the equilibrium force, $f_\mathrm{eq}$, and of the maximum of the dynamical force, $f_\mathrm{m}$, as a function of $d$ normalized by $e$.
\label{fig_dynam}}
\end{figure}

Two striking behaviors were observed in Ref.~\cite{Fournier14} (see Fig.~\ref{fig_dynam}): (i) the force $f(t)$ reaches a maximum $f_\mathrm{m}$ and then decreases to the equilibrium force $f_\mathrm{eq}$. (ii) While $f_\mathrm{eq}$ decreases exponentially with the separation $d$ between the inclusions, the maximum force $f_\mathrm{m}$ decreases as a power-law $\sim d^{-3}$ until it reaches $f_\mathrm{eq}$. Hence $f_\mathrm{m}$ is long-ranged. Although these results were obtained with a simplified Hamiltonian for the inclusions, it is likely that the general trends observed will also apply to more realistic cases. It should be straightforward to extend the model of Ref.~\cite{Fournier14} to inclusions that trigger at different times, but considering the movement of the inclusions at the same time as the movement of the membrane would be more challenging.

\subsection{Other geometries}

Until now, we focused on the case of inclusions with separation $d$ larger than their characteristic size, embedded in a membrane with small deformations around the flat shape. This is the case that has attracted the most attention in the literature, because of its relevance for proteins embedded in the membrane, and because of its technical tractability. We now move on to other geometries.

\subsubsection{Spherical vesicle}

Ref.~\cite{Dommersnes98} focused on the membrane-mediated interaction arising from the ground-state deformation between two disk-shaped inclusions embedded in the membrane of a spherical vesicle, and imposing contact angles. The case of the spherical vesicle is practically relevant both in biology and in in-vitro experiments. The energy of the membrane was considered to be dominated by bending rigidity, which requires the length scales at play (in particular the vesicle radius) to be small with respect to $\sqrt{\kappa/\sigma}$. The covariant Helfrich Hamiltonian (Eq.~\ref{HelfCov} with no Gaussian curvature term) was adapted to small deformations with respect to a sphere. 

The interaction was evaluated thanks to an expansion of the energy-minimizing profile of the membrane, and it was found to be strongly enhanced with respect to the flat-geometry interaction (Eq.~\ref{ElastGou}) at length scales where the spherical shape of the vesicle is relevant. At sufficient angular separation, the effective power law of the interaction is $\sim1/d^{1/3}$~\cite{Dommersnes98}. This sheds light on the strong impact of the underlying geometry of the membrane on membrane-mediated forces. Qualitatively, in a flat membrane, the interaction is weaker because the curvature energy in Eq.~\ref{HelfCov} can be minimized quite well between the inclusions (with an almost perfect saddle that has very little curvature energy), which is not possible in the spherical geometry. Similarly, in the case of external torques (Sec.~\ref{extft}), the imposed orientations did not allow for this low-energy saddle, thus enhancing the interaction.

\subsubsection{Close parallel rods}
\label{closeRods}

We already discussed the case of rigid rods of length $L$, at a distance $d\gg L$~\cite{Golestanian96a, Golestanian96b}, which is close to the point-like case. The opposite regime $d\ll L$ is also relevant biologically, since it can model semi-flexible polymers adsorbed on the membrane. In Ref.~\cite{Podgornik95}, the effect of the reduction of the membrane fluctuations by the presence of a semiflexible (wormlike) polymer was discussed. An effective nematic interaction was found between different segments of the polymer, and it was shown that this interaction can yield an orientational ordering transition.

\begin{figure}[htb] 
\centering
\includegraphics[width=\textwidth]{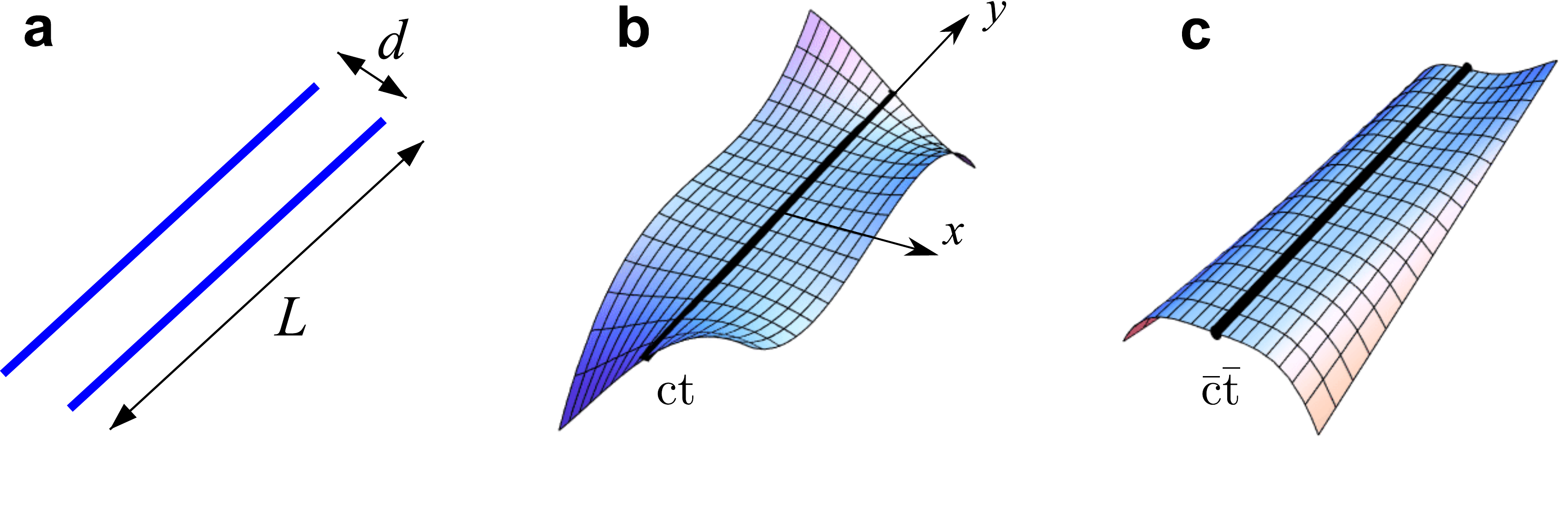}
\caption{Rods embedded in membranes. (a) Geometry: two parallel rods of length $L$ at separation $d\ll L$. (b) and (c) Two examples of rod types.  All rigid rods impose a vanishing curvature along them: $\partial_y\partial_y h=0$ on the rod. (b) Rod that allows curving (``c'') and twisting (``t'') across it. (c) Rod that does not allow curving or twisting across it: it imposes $\partial_x\partial_x h=0$ and $\partial_x\partial_y h=0$ as well as $\partial_y\partial_y h=0$ (see Ref.~\cite{Bitbol11_rods}). \label{Schema_batons}}
\end{figure}

Let us first consider rods that do not break the up-down symmetry of the membrane. The case of such stiff parallel rods in the limit $d\ll L$ (see Fig.~\ref{Schema_batons}a) embedded in a membrane with energy dominated either by bending rigidity (Eq.~\ref{HelfSansSigma}) or by tension (Eq.~\ref{HelfSltSigma}) was studied in Ref.~\cite{Golestanian96c}. A constant scale-free Casimir-like interaction per unit length is then expected~\cite{Li92}, and indeed the Casimir-like interaction potential is then proportional to $-k_\mathrm{B}T\, L/d$~\cite{Golestanian96c}. This interaction is much stronger than the one between point-like objects (Eq.~\ref{CasiGou}), because the constraints imposed on fluctuation modes are much stronger in the geometry of parallel close rods. Ref.~\cite{Golestanian96c} further showed that such rods tend to bend toward one another below a certain critical distance, and that their interaction is screened by out-of-plane fluctuations if the rigidity of the polymer is finite.

This situation was further studied in Ref.~\cite{Bitbol11_rods}. Rods were modeled as constraints imposed on the membrane curvature along a straight line, allowing to define four types of rods, according to whether the membrane can twist along the rod and/or curve across it (see Fig.~\ref{Schema_batons}b-c for two examples of these rod types). The numerical prefactors of the potential in $L/d$ were obtained for interactions between the different types of rods, and they were all found to be attractive, provided that the rods are rigid, i.e. that they impose $\partial_y\partial_y h=0$ along them, with the notations of Fig.~\ref{Schema_batons}. However,  repulsion was obtained between objects imposing completely antagonistic conditions (i.e. a rigid rod only imposing $\partial_y\partial_y h=0$ along it, see Fig.~\ref{Schema_batons}b, and a non-rigid ``ribbon'' only imposing $\partial_x\partial_x h=0$ along it), which is reminiscent of results obtained in critical binary mixtures~\cite{Gambassi09}. In addition, the interaction energy was studied numerically versus $d/L$, thanks to a discretization scheme~\cite{SinRonia12}, showing the transition between the asymptotic behaviors at large $d/L$~\cite{Golestanian96a} and at small $d/L$~\cite{Bitbol11_rods} were recovered. Finally, the bending and coming into contact of the rods due to the fluctuation-induced interaction was discussed: it was predicted to occur below a certain value of $d$~\cite{Bitbol11_rods}. 

The $L\gg d$ geometry gives insight into what happens between two generic inclusions that are very close to one another, through the proximity force approximation~\cite{Derjaguin34}. This approximation was used in the case of disk-shaped inclusions in Refs.~\cite{Lin11, Bitbol11_rods}, showing that the fluctuation-induced interaction potential then scales as $1/d^{1/2}$.

In Ref.~\cite{Weikl03}, the interaction due to the ground-state deformation between parallel rigid cylinders adsorbed on a membrane and interacting with it through an adhesion energy was studied. The membrane was assumed to be in the regime of small deformations, but both tension and bending were accounted for (see Eq.~\ref{HelfAvecSigma}), and the geometry where $d\ll L$ was considered. The interaction due to the ground-state deformation was calculated explicitly in this effectively one-dimensional case. It was found to be repulsive for a pair of cylinders adhering to the same side of the membrane, and attractive for cylinders adhering to opposite sides (and hence imposing an opposite curvature). This is at variance with the point-like case, where the interaction only depends on the modulus of the curvatures imposed (see Eq.~\ref{ElastGou}). The dependence in $d$ is in $\tanh(d/\sqrt{\kappa/\sigma})$ in the first case, and in $\coth(d/\sqrt{\kappa/\sigma})$ in the second one~\cite{Weikl03}.

\subsubsection{Large deformation regime}

All cases discussed until now focused on small deformations. Then, the Hamiltonian of the membrane is quadratic, and the field theory is Gaussian. This provides tractability, both to solve the Euler-Lagrange equations that give the ground-state shape, which are then linear, and to compute thermodynamical quantities such as the free energy. Here, we will discuss the biologically relevant but much trickier regime of large deformations.

In Ref.~\cite{Muller05, Muller05b}, the covariant membrane stress and torque tensors associated to the full Helfrich Hamiltonian~\cite{Capovilla02} were used to determine formal expressions of the forces between objects adsorbed on fluid membranes that are due to the ground-state deformation of the membrane. These expressions are valid without assuming small deformations, but the ground-state shape needs to be determined in order to obtain a more explicit expression. This is not an easy task in the large-deformation regime. Equilibrium shapes in the large deformation regime were further investigated in Ref.~\cite{Muller07}, allowing to plot the force between cylinders, in the case of a fixed adhesion area between them and the membrane. The direction of the force and its asymptotic exponential decay at large $d/\sqrt{\kappa/\sigma}$ were found to remain the same as in the small-deformation regime~\cite{Weikl03}. This situation was also investigated numerically in Ref.~\cite{Mkrtchyan10} in the case of cylinders interacting with the membrane through an adhesion energy, yielding phase diagrams of the system.

In Ref.~\cite{Gosselin11}, the entropic contribution to the membrane-mediated interaction between two long cylinders adsorbed on the same side of a membrane was studied in the regime of large deformations, in the case of a fixed adhesion area between the cylinders and the membrane. The free energy of the system was calculated by assuming Gaussian fluctuations around the ground-state shape. Interestingly, this entropic contribution enhances the ground-state repulsion between the two cylinders~\cite{Gosselin11}, while the fluctuation-induced interaction between identical rods in the small-deformation regime is attractive~\cite{Golestanian96c,Bitbol11_rods}. This is presumably a non-trivial effect coming from the non-linearities at play in the large deformations. It would be interesting to go beyond the approximation of Gaussian fluctuations around the ground-state shape.

\begin{figure}[htb] 
\centering
\includegraphics[width=0.85\textwidth]{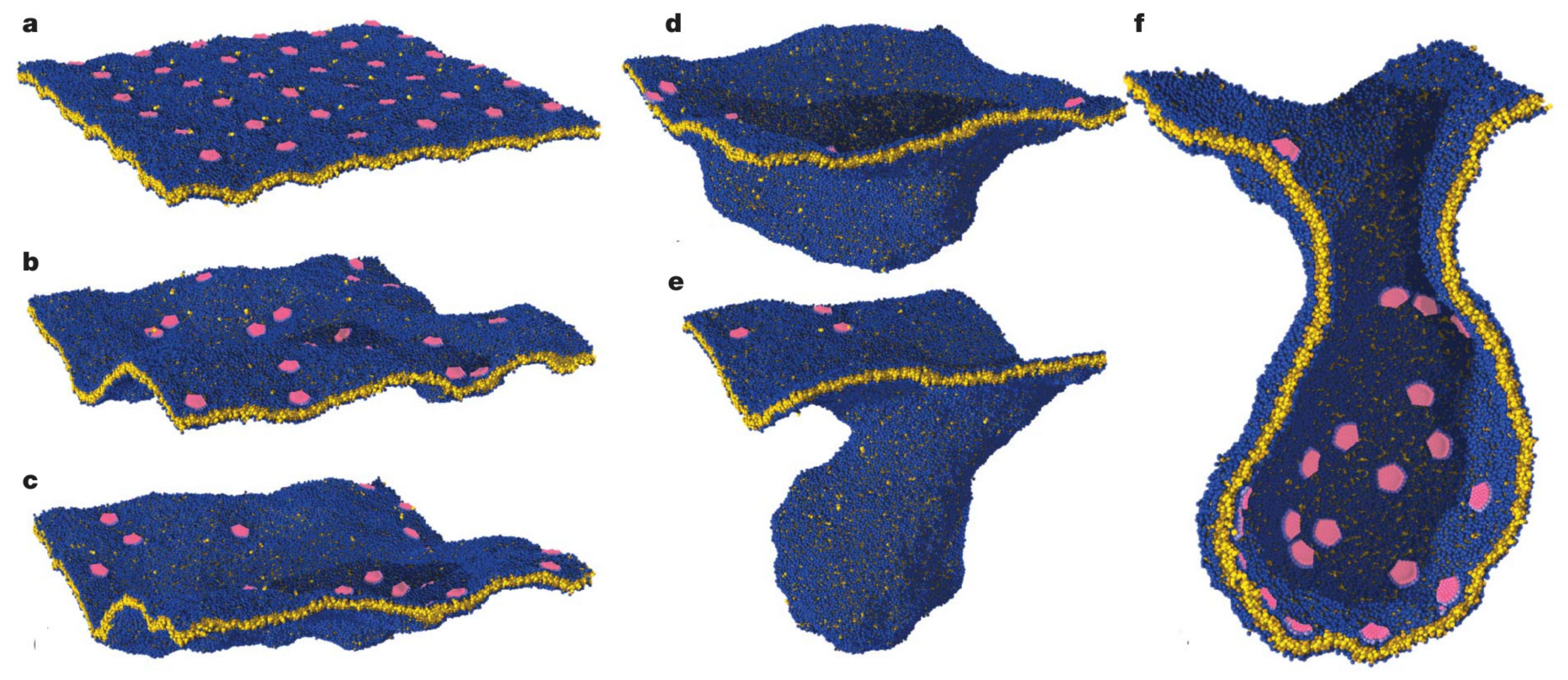}
\caption{Successive snapshots of a coarse-grained simulation of a membrane with several curvature-inducing inclusions. A process of vesiculation is induced by the elastic interaction between inclusions, which becomes attractive at short separations. Reproduced from Ref.~\cite{Reynwar07}. \label{Vesiculation}}
\end{figure}

Solving the shape/Euler-Lagrange equation for membranes beyond the domain of small deformations is technically very hard for most geometries, and incorporating fluctuations too, but numerical simulations can provide further insight. The coarse-grained molecular-dynamics membrane simulations without explicit solvent description of Ref.~\cite{Reynwar07} showed that the elastic interaction between two isotropic curvature-inducing membrane inclusions (quasi-spherical caps) can become attractive at short separations, provided that the inclusions induce a strong enough curvature. Recall that the interaction due to the ground-state deformation, which is dominant with respect to the fluctuation-induced one for large enough curvatures imposed by inclusions, is always repulsive in the regime of small deformations (see Eq.~\ref{ElastGou}). This hints at highly non-trivial effects of the large-deformation regime. The attractive membrane-mediated interaction was found to be able to yield aggregation of the caps and vesiculation of the membrane~\cite{Reynwar07} (see Fig.~\ref{Vesiculation}). The case of curved phase-separated lipid domains was explored in Ref.~\cite{Yuan11} through coarse-grained molecular-dynamics simulations. The interaction between domains was found to be attractive, but the angles imposed by the domains were smaller than those yielding attraction in Ref.~\cite{Reynwar07}.

A numerical minimization via Surface Evolver of the Helfrich Hamiltonian Eq.~\ref{HelfCov} for a membrane with two in-plane isotropic curvature-inducing inclusions was presented in Ref.~\cite{Reynwar11}, and forces were calculated by studying infinitesimal displacements. A change of sign of the membrane-mediated interaction due to the ground-state deformation of the membrane was obtained, consistently with Ref.~\cite{Reynwar07}. The repulsive interaction, agreeing quantitatively with Eq.~\ref{ElastGou} at large $d/a$ and for small deformations, turned attractive for $d/a$ of order one, provided that the curvature imposed by the inclusions (and hence the membrane deformation) was large enough. The separation $d$ is defined as the center-to-center distance projected on a reference plane, while $a$ is the real radius of the inclusions, so that in the large deformation regime where inclusions are very tilted, it is possible to have $d<2a$. Attraction occurs in this regime, which is inaccessible to the small-deformation approach. Recently, Ref.~\cite{Schweitzer15} studied anisotropic protein scaffolds, modeling e.g. BAR proteins, in the large-deformation regime, through similar numerical minimization methods: strongly anisotropic attractive interactions were obtained.

Ref.~\cite{Bahrami12} presented a Monte-Carlo simulation of spherical nanoparticles adsorbed on a spherical vesicle modeled as a triangulated surface. Aggregation of the nanoparticles and inward tubulation of the vesicle were observed, implying strong attractive interactions. Note however that adhesion might have a strong impact on these structures~\cite{Saric13}. A similar coarse-grained description of a membrane vesicle was used in Ref.~\cite{Ramakrishnan13} to investigate the collective effects of anisotropic curvature-inducing inclusions, modeling e.g. BAR proteins. Vesicles were strongly deformed by the numerous inclusions, with sheet-like shapes or tubulation depending on inclusion concentration, and aggregation and nematic ordering of these inclusions were observed.

\subsection{Experimental studies}

While membrane-mediated interactions have been the object of significant theoretical and numerical attention, quantitative experimental tests of the theoretical predictions remain scarce to this day. A very active research area in biophysics deals with the morphological changes of the cell (invagination \cite{Roemer07}, vesiculation \cite{Reynawar07}, etc.) under the action of various proteins (see \cite{McMahon05} for a recent review). However, many other ingredients than membranes and inclusions are at play in these biological systems, for instance the cytoskeleton, out-of-equilibrium events, etc., which makes it hard to isolate membrane-mediated interactions. Biomimetic lipid membranes such as giant unilamellar vesicles~\cite{Angelova86} are a good model system to study such effects. In principle, the inclusions could be real proteins, but these molecules have complex shapes, which makes it difficult to test predictions of models developed for simple geometries. Many studies have focused on the simpler and more easily controlled system of colloids adhering to membranes (see Ref.~\cite{Saric13} for a review), and some have investigated interactions between phase-separated membrane domains~\cite{Semrau09}. However, even in these simpler cases, membrane-mediated interactions may involve other effects, such as adhesion of the colloids, variability of contact angles imposed by domains, etc.

An experimental study of the aggregation of spherical colloidal particles adhering to biomimetic lipid membranes was presented in Ref.~\cite{Koltover99}. The observed aggregation of two particles was deemed consistent with a short-range (e.g. exponential) attractive force, and no signature of a longer-range force was obtained. Note that theoretical studies predict a mostly repulsive membrane-mediated force in this geometry, except at very high deformations and small distances. Surprisingly, triplets were observed to form chains, and a linear ring-like aggregate was observed around the waist of a vesicle. Linear chain-like arrangements were also obtained in simulations of a very similar situation in Refs.~\cite{Yue12,Saric12}, for certain sizes of particles and adhesion regimes. Ref.~\cite{Saric12} used a scaling argument to show that this was not due to membrane-mediated interactions, but to the adhesion of the particles to the membrane, as a linear aggregate yields a higher adhesion area than a compact one. Such a phenomenon would thus not arise in the case of inclusions~\cite{Saric13}. 

\begin{figure}
	\centerline{\includegraphics[width=0.8\textwidth,angle=0]{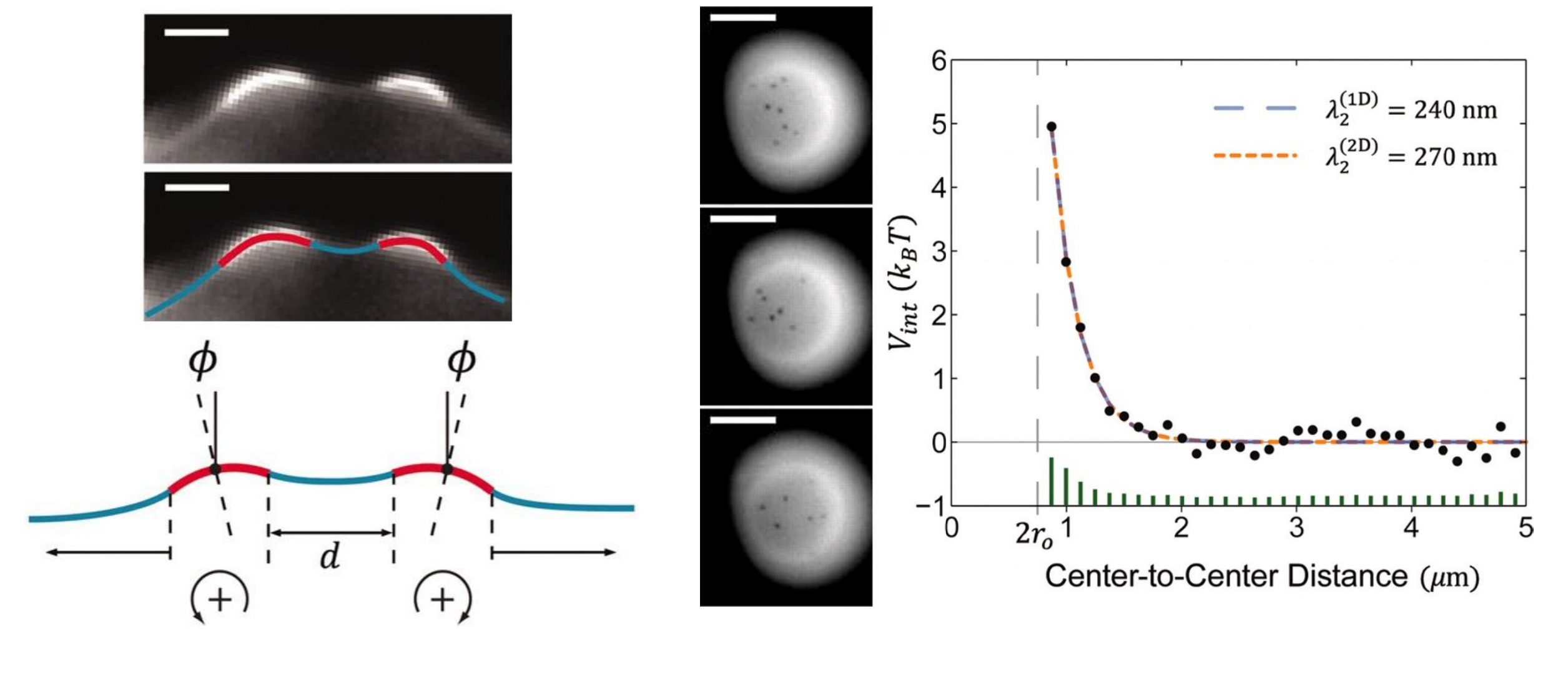}}%
	\caption{\label{fig:Ursell09}Shape of the dimpled domains (left), interacting domains on the surface of the same vesicle (center) and repulsive interaction potential, with a fit to theoretical predictions from Ref.~\cite{Weikl98} (right). Adapted from Figures~3 and 4 of reference \cite{Ursell09}.}%
\end{figure}

Apart from proteins and colloids, another source of membrane deformation is the presence of phase separated (liquid-ordered/liquid disordered) domains, which can be partially budded. Contrast between the domains is obtained in fluorescence microscopy by adding a dye which partitions into one phase~\cite{Rozovsky05} or by selectively labeling one lipid species~\cite{Ursell09}. Selective deuteration can also be used to induce contrast in small-angle nuclear scattering \cite{Heberle13}. In Ref.~\cite{Semrau09}, the stability of partially budded  domains was interpreted as a signature of repulsive interactions, since flat ones rapidly fused. The strength of this interaction was evaluated by measuring the distribution of inter-domain distance, and then by evaluating the effective spring constant of the confining potential. It was found to be consistent with the membrane-mediated interaction arising from the ground-state deformation of a tension-free membrane in the small-deformation and large-separation regime (Eq.~\ref{ElastGou}). In Ref.~\cite{Ursell09}, a good agreement was obtained between the observed in-plane distribution of the domains and the predictions of the elastic theory in the presence of tension~\cite{Weikl98} (see Figure~\ref{fig:Ursell09}). 

\section{Short-range membrane-mediated interactions}
\label{sec:short}

In Part 1, we dealt with long-range membrane-mediated interactions between inclusions, which arise from the curvature constraints imposed by rigid inclusions. There exist several other ways in which inclusions can couple to the surrounding membrane and thus interact with other inclusions through the membrane, but these effects are generally short-ranged. The study of these interactions was in fact initiated before that of their long-range counterparts~\cite{Goulian93etc}. Membrane proteins were shown experimentally to tend to immobilize neighboring lipids~\cite{Jost73}. A membrane-mediated attraction between proteins was predicted to arise due to this local ordering~\cite{Marcelja76}, and to decay exponentially above the correlation length of the membrane order parameter~\cite{Schroder77}. Proteins can locally perturb the thickness of the membrane due to this local ordering, but they may also couple preferentially to one component of a lipid mixture~\cite{Owicki78}. 

Here, we are going to focus on the coupling of proteins to membrane thickness. Intrinsic membrane proteins can have a \emph{hydrophobic mismatch} with the membrane: their hydrophobic thickness is slightly different from that of the unperturbed membrane. Hydrophobic mismatch is ubiquitous, and has important biological consequences, since the activity of many membrane proteins has been shown to depend on membrane thickness~\cite{Killian98}. As proteins are more rigid than membranes, the membrane generically deforms in the vicinity of the protein, in order to match its thickness and avoid exposing part of the hydrophobic chains of lipids to water. This local deformation of the membrane thickness yields a membrane-mediated interaction between two such proteins. 

Membrane thickness deformations are not included in the traditional Helfrich description of the membrane~\cite{Helfrich73}. Describing them is tricky since they occur on the nanometer scale, which corresponds to the limit of validity of usual continuum theories where only long-scale terms are kept. Let us focus on these models before moving on to the actual interactions.

\subsection{Models for local membrane thickness deformations}
\label{earlymodels}
\subsubsection{Early models}

The idea that the membrane hydrophobic thickness must locally match that of an intrinsic protein was first used in theoretical descriptions of lipid-protein interactions that focused on the thermodynamic phase behavior of the lipid-protein system and on protein aggregation. In Ref.~\cite{Mouritsen84}, a thermodynamic model called the ``mattress model'' was proposed in order to describe the phase diagrams of lipid bilayers containing proteins with a hydrophobic mismatch. 

More detailed theoretical investigations of local membrane thickness deformations and of resulting membrane-mediated interactions were motivated by experimental results on the antimicrobial peptide gramicidin. In lipid membranes, two gramicidin monomers, one on each side of the bilayer, can associate to form a dimer, which acts as an ion channel. While isolated monomers do not deform the membrane, the dimeric channel generically possesses a hydrophobic mismatch with the membrane~\cite{Kelkar07}. Conductivity measurements yield the formation rate and lifetime of the channel, which are directly influenced by membrane properties~\cite{Kolb77,Elliott83,Goulian98}. Hence, gramicidin constitutes a very convenient experimental system to probe the effects of local membrane thickness deformations.

The first attempt to explain the dependence of gramicidin channel lifetime on the membrane thickness was provided by Ref.~\cite{Elliott83}. It is based on the idea that the relevant membrane energy variation upon dimer breaking is mostly due to membrane tension, which pulls apart the monomers in a membrane with hydrophobic thickness larger than that of the dimer. The resulting estimate of the gap between the two monomers in the transition state is $\delta\simeq 1.8$~nm~\cite{Elliott83}. However, this is far larger than the separation required for the breaking of the hydrogen bonds that stabilize the dimer~\cite{Kelkar07}, which is of order 1~\AA. Hence, this first model was not complete.

\subsubsection{Huang's model}

\begin{figure}[htb] 
\centering
\includegraphics[width=0.45\textwidth]{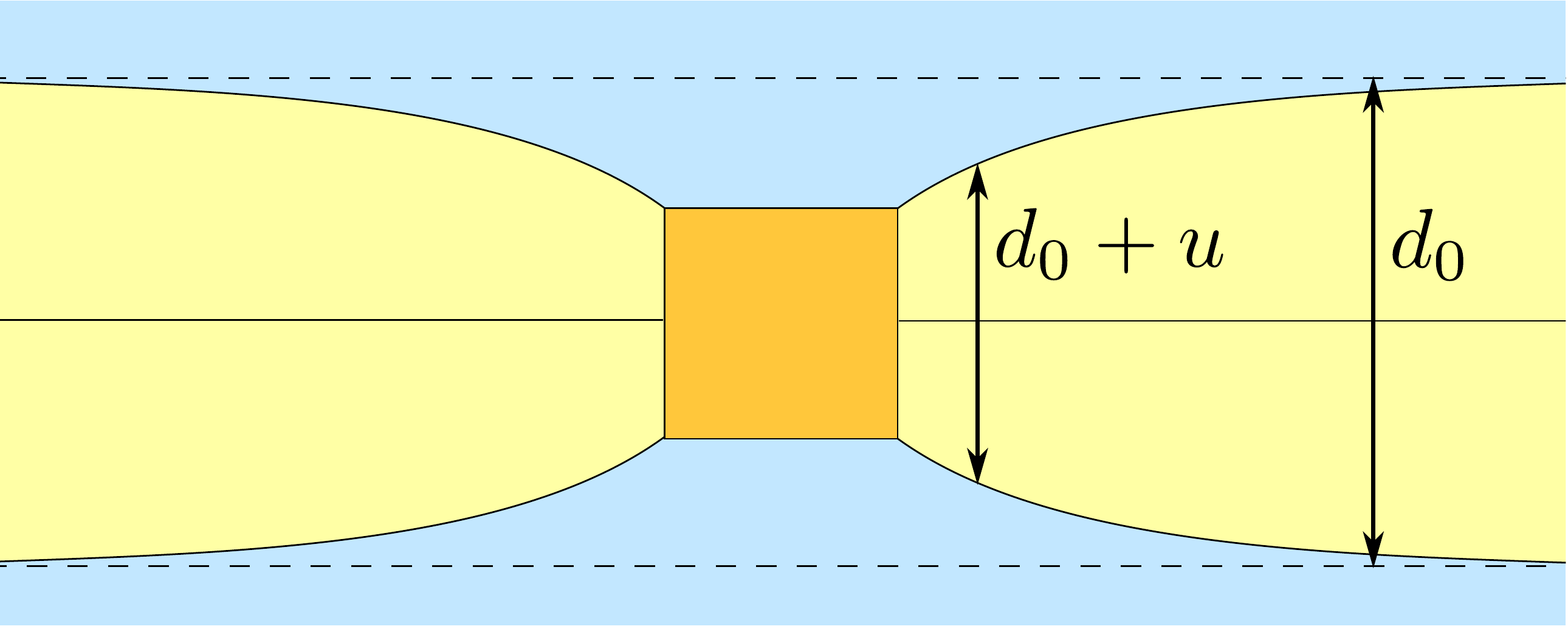}
\caption{Cut of a bilayer membrane (yellow) containing a protein with a hydrophobic mismatch, represented as a square (orange). The equilibrium thickness of the bilayer is $d_0$, while the actual thickness is denoted by $d_0+u$.\label{d0u}}
\end{figure}

The first full continuum model describing membrane thickness deformations was proposed in Ref.~\cite{Huang86}. The Hamiltonian per unit area of the membrane was written by analogy with a smectic A liquid crystal, in which the elongated molecules organize in layers with the molecules oriented along the layers' normal. These two systems present the same symmetries. The most important energetic terms in smectic A liquid crystals correspond to compression of the layers, and to splay distortion, i.e. curvature orthogonal to the layers~\cite{dG_LC}. In addition, the contribution of the ``surface tension'' of the membrane was included~\cite{Huang86}. Restricting to symmetric deformations of the two monolayers, the effective Hamiltonian $H$ of the membrane reads~\cite{Huang86}
\begin{equation}
H=\int dxdy\left[\frac{K_a}{2\,d_0^2}\,u^2+\frac{\gamma}{4}\,(\bm{\nabla}u)^2+\frac{\kappa}{8}\,\left(\nabla^2 u\right)^2\right]\,.
\label{Huang}
\end{equation}
In this expression, $u$ denotes the thickness excess of the membrane relative to its equilibrium thickness $d_0$ (see Fig.~\ref{d0u}), $K_a$ is the stretching modulus of the membrane, $d_0$ its equilibrium thickness, $\gamma$ its ``surface tension'', and $\kappa$ an elastic constant associated to splay. Finally, $x$ and $y$ denote Cartesian coordinates on the mid-plane of the membrane.

Ref.~\cite{Huang86} assimilated $\gamma$ to the tension of a Plateau border and $\kappa$ to the Helfrich bending modulus, which may be questioned (see below). The corresponding typical values allowed to neglect the contribution of the ``tension'' term. By minimizing the resulting membrane Hamiltonian, analytical expressions were obtained for the membrane deformation profiles close to a mismatched protein such as the gramicidin channel, obtaining a decay length of a few nanometers. This model yields a satisfactory agreement with the experimental results of Ref.~\cite{Elliott83}.

\subsubsection{Models based on the work of Dan, Pincus and Safran}
\label{DPS}

Refs.~\cite{Dan93,ArandaEspinoza96} proposed another construction of the membrane Hamiltonian associated to thickness deformations. The energy per lipid molecule in each monolayer of the membrane was written for small deformations as a generic second-order expansion in the variation of area per lipid and in the local ``curvature'' of the monolayer thickness (different from the curvature of the shape of the membrane involved in the Helfrich model, which disregards thickness). 
Incompressibility of the lipids was used to relate the monolayer thickness and and the area per lipid. Using the same notations as in Eq.~\ref{Huang}, and restricting again to up-down symmetric deformations of the membrane, the membrane Hamiltonian of Ref.~\cite{ArandaEspinoza96} reads:
\begin{equation}
H=\int dxdy\left[ \frac{K_a}{2\,d_0^2}u^2 +\frac{\kappa \,c_0}{2}\,\nabla^2 u+\frac{\kappa}{2\,d_0}\left(c_0-c'_0\Sigma_0\right)u\,\nabla^2 u+\frac{\kappa}{8}\left(\nabla^2 u\right)^2\right]\,,\label{safr}
\end{equation}
where $c_0$ is the spontaneous curvature of a monolayer, while $c'_0$ denotes its derivative with respect to the area per molecule, and $\Sigma_0$ the equilibrium area per lipid. 

The main difference between this model and that of Ref.~\cite{Huang86} is that the effect of monolayer spontaneous curvature is included in Eq.~\ref{safr}. It was shown in Refs.~\cite{Dan93, ArandaEspinoza96} that this ingredient can yield oscillations in the membrane deformation profile, and in the resulting interaction potential between two mismatched proteins. Note that no ``tension'' term is included in Eq.~\ref{safr}, but the ``tension'' term in Eq.~\ref{Huang} was neglected in all the calculations of Ref.~\cite{Huang86} too.

The model of Refs.~\cite{Dan93, ArandaEspinoza96} was generalized in Refs.~\cite{Brannigan06, Brannigan07}, where results of coarse-grained molecular-dynamics simulations for mismatched proteins in lipid membranes were also presented. The deformations of the average shape of the membrane (i.e., those usually described by the Helfrich model), and the small-scale protrusions were accounted for, as well as the symmetric thickness deformations~\cite{Brannigan06, Brannigan07}. The effect of Gaussian curvature was also included in Ref.~\cite{Brannigan07}, and Ref.~\cite{Watson11} added the effect of tilt.

The model of Refs.~\cite{Dan93, ArandaEspinoza96,Brannigan06, Brannigan07} was further generalized in Ref.~\cite{Bitbol12_grami}, where an additional term, proportional to the squared gradient of thickness, was included in the initial expression of the energy per lipid molecule in each monolayer of the membrane. Physically, this term should involve a microscopic interfacial tension contribution, associated to variations of the area per lipid. Note that this is different from the Plateau border tension discussed and discarded in Ref.~\cite{Huang86}, as in a Plateau border, molecules can move along the surface and exchange with the bulk, yielding a smaller tension. Macroscopic membrane tension was also incorporated explicitly in Ref.~\cite{Bitbol12_grami}, through a chemical potential $\mu$ set by the rest of the membrane on the patch considered: $\sigma=-2\mu/\Sigma_0$ then plays the part of an externally applied tension. The Helfrich Hamiltonian with tension Eq.~\ref{HelfAvecSigma} was recovered from this model for average height deformations. In the case where the average shape of the membrane is flat, and integrating out anti-symmetric thickness deformations to focus on symmetric ones, the Hamiltonian reads:
\begin{equation}
H=\int dx dy\left[\frac{\sigma}{d_0}\,u+\frac{K_a}{2\,d_0^2}\,u^2+\frac{K'_a}{2}\,(\bm{\nabla}u)^2+\frac{K''_a}{2}\,(\nabla^2 u)^2\right]\,,
\label{jbarc2b}
\end{equation}
plus omitted boundary terms (see Ref.~\cite{Bitbol12_grami}), with
\begin{align}
K'_a&=-\frac{\kappa_0}{d_0}(c_0-c'_0\Sigma_0)+k'_a\,,\label{Kpa}\\
K''_a&=\frac{\kappa_0}{4}\,,\label{Ksa}
\end{align}
and the same notations as in Eqs.~\ref{Huang} and~\ref{safr}, and where the new contribution $k'_a$ with respect to Eq.~\ref{safr} arises from the term proportional to the squared gradient of the thickness $u$. (The definition of $u$ in Ref.~\cite{Bitbol12_grami} is slightly different from that of Refs.~\cite{Dan93, ArandaEspinoza96,Brannigan06, Brannigan07}, but it does not affect the present discussion.)

The predictions of the model of Ref.~\cite{Bitbol12_grami} were compared with numerical profiles of membrane thickness close to a mismatched protein~\cite{Brannigan06, Brannigan07, West09}, and with experimental data regarding gramicidin lifetime~\cite{Elliott83} and formation rate~\cite{Goulian98}. This analysis yielded consistent results for the term stemming from the gradient of the area per molecule, and its order of magnitude was found to be of order of the contribution of the interfacial tension between water and the hydrophobic part of the membrane. In addition, the presence of this new term allowed to explain for the first time a systematic dependence in previous numerical data.

\subsubsection{Isotropic cross-section}
\label{sr_iso}

The first models of short-range interactions between transmembrane proteins assumed that the proteins are coupled to a local order parameter describing the internal state of the membrane,  either the conformational/chain-packing properties of the lipids, or the bilayer thickness $u$~\cite{Marcelja76,Schroder77}. Both are equivalent for a fully incompressible membrane hydrophobic core. In Refs.~\cite{Owicki78,Owicki79}, a generic Landau--Ginzburg expansion of the free energy density in terms of $u$ and its first gradient was used to investigate the energy of a hexagonal lattice of embedded proteins imposing a value $u_0$ of the order parameter, i.e., a fixed hydrophobic mismatch, on their edge. Approximating the Wigner-Seitz cell of the lattice by a circle, which yields cylindrical symmetry, the authors derived a monotonically attractive short-range interaction caused by the overlap of the membrane regions deformed by the inclusions.

As discussed in Sec.~\ref{earlymodels}, several models based on the thickness order parameter $u$ have been developed. They have been used to study membrane-mediated interactions. These models essentially introduced terms involving the second-order derivative of $u$, based on the (recently questioned~\cite{Bitbol12_grami}) expectation that the term proportional to $(\bm{\nabla} u)^2$ was negligible.
In particular, Ref.~\cite{Huang86} introduced a term proportional to $(\nabla^2u)^2$, by analogy with the splay term for smectic liquid crystals. Later, Ref.~\cite{Dan93} introduced additional terms, linear in $\nabla^2u$ and in $u\nabla^2u$, which arise from the spontaneous curvature of the monolayers and its dependence on the area per lipid. This initiated a series of works~\cite{Dan93,Dan94,ArandaEspinoza96} aiming to estimate the elastic energy of a hexagonal lattice of proteins with hydrophobic mismatch. These works showed that the interaction potential can be non-monotonous, with short-distance repulsion and a minimum energy at finite separation. These effects can arise from the spontaneous curvature term, but also from a fixed contact angle between the membrane hydrophobic-hydrophilic interface and the inclusion, thereafter referred  to as ``slope". 
The associated multi-body effects were investigated in Ref.~\cite{Harroun99_theo} through a Monte-Carlo simulation of inclusions fixing both the membrane thickness and its slope, in a membrane described by the elastic energy in Eq.~\ref{Huang}. This study also demonstrated the interest of the structure factor to test the models. Another term involving second-order derivatives of the thickness profile $u$, proportional to its Gaussian curvature, was included in Ref.~\cite{Brannigan07}, improving the agreement with coarse-grained molecular-dynamics numerical simulations. Note that oscillations in the interaction potential were observed in the coarse-grained molecular-dynamics simulations of Ref.~\cite{Neder11}.

The term proportional to $(\bm{\nabla} u)^2$ in the elastic energy density was originally discarded on the grounds that it  originates from a negligible microscopic surface tension assimilated to that of a Plateau border~\cite{Huang86}. However, it was recently shown by us to also originate form gradients of lipid density, and therefore to contribute significantly to the elastic Hamiltonian~\cite{Bitbol12_grami}. Note in addition that the term in $u\nabla^2u$ introduced in Refs.~\cite{Dan93,ArandaEspinoza96} contributes to the $(\bm{\nabla} u)^2$ term once integrated by parts.
 
In the end, these models converge towards the most general quadratic expansion in terms of $u$ and its first and second-order derivatives~\cite{Bitbol12_grami,Bories_tbp}. In standard statistical field theory, it is justified to neglect higher-order gradients, because the focus is on large-scale physics and the coarse-graining length is much larger than the range of the microscopic interactions~\cite{ChaikinLubensky}. However, here, such arguments do not hold since the distortions around proteins relax on a length comparable with the bilayer thickness. Therefore, in practice, one should rather rely on comparison with experiments and simulations to determine how many terms to include in the expansion. Our current understanding is that all linear and quadratic terms involving derivatives of $u$ up to second order should contribute, and that the best strategy is to try to fit the parameters of the elastic Hamiltonian and of the protein--membrane coupling using experimental or numerical data~\cite{Bories_tbp}.

The focus of this chapter is on membrane-mediated interactions arising from direct constraints on the membrane shape (mean shape and thickness). Hence, we will not discuss in detail the role of the underlying lipid tilt degree of freedom~\cite{Watson11} in membrane-mediated interactions~\cite{Fournier98,May99,May00,May02,Bohinc03,Kozlovsky04}. However, tilt certainly plays a part in these interactions. For instance, proteins with no hydrophobic mismatch but with a hour-glass shape~\cite{Fournier98,May99} may induce a membrane deformation due to the boundary conditions they impose on lipid tilt. A legitimate question, though, is how necessary it is to include this degree of freedom. Statistical physics allows to integrate out virtually any degree of freedom~\cite{ChaikinLubensky}. The resulting effective elasticity for the remaining degrees of freedom takes into account the underlying distortion energy of the removed ones. For instance, integrating out the tilt degree of freedom in the presence of an hour-glass shaped inclusion would produce an effective boundary energy depending on the inclusion thickness and on its angle with the membrane. What is not clear is how many orders in the derivatives of $u$, both in the bulk and in the boundary energy, one would have to introduce in order to properly account for the removed degrees of freedom. Future works in this direction could be interesting.

\subsubsection{Anisotropic cross-section}

While most theoretical studies of short-range membrane-mediated interactions have considered cylinder-shaped inclusions, actual membrane proteins have various shapes. As in the case of long-range interactions, in-plane anisotropy may result in directional membrane-mediated interactions, which may impact the formation of multi-protein complexes. 

In Ref.~\cite{Haselwandter13}, an analytical method was developed to study membrane-mediated interactions between in-plane anisotropic mismatched inclusions. The effective Hamiltonian $H$ associated to membrane thickness deformations was expressed as
\begin{equation}
H=\int dxdy\left\{\frac{K_a}{2 d_0^2}\,u^2+\gamma\left[\frac{u}{d_0}+\frac{(\bm{\nabla}u)^2}{8}\right]+\frac{\kappa}{8}\,\left(\nabla^2 u\right)^2\right\}\,,\label{Hasel}
\end{equation}
where we have used the notations defined in Eq.~\ref{Huang}. This model is based on that of Ref.~\cite{Huang86} (see Eq.~\ref{Huang}), but includes an additional ``tension'' term in $u/d_0$. Such a term is also included in Ref.~\cite{Bitbol12_grami} (see Eq.~\ref{jbarc2b}), but without the assumption that its prefactor is related to that of the squared thickness gradient term. This assumption should be viewed as a simplifying hypothesis, given the contribution of monolayer curvature to the squared thickness gradient term~\cite{Dan93, ArandaEspinoza96,Brannigan06, Brannigan07} and the difference between externally applied tension and interfacial tension~\cite{Bitbol12_grami} (see Sec.~\ref{DPS}). 

In Ref.~\cite{Haselwandter13}, the solution of the Euler-Lagrange equation associated with Eq.~\ref{Hasel} in the case of a single cylinder-shaped inclusion was expressed using Fourier-Bessel series. Then, using an ansatz introduced in Ref.~\cite{Weikl98} in the context of long-range membrane-mediated interactions, the ground-state shape of the membrane in the presence of two inclusions was written as a sum of two such series. The coefficients of the successive terms of these series can be chosen in order to match the boundary conditions imposed by both inclusions, using expansions in $a/d<1$.

This method was extended to weakly anisotropic inclusions, modeling mechanosensitive channels of large conductance (MscL) in Ref.~\cite{Haselwandter13}. The in-plane cross-section of these pentameric proteins was described as a circle perturbed by a small-amplitude sinusoidal, with fifth-oder symmetry. Boundary conditions along the edge of these proteins were expressed perturbatively in the amplitude of the sinusoidal, allowing to use the method described above. The resulting anisotropic membrane-mediated interaction features an energy barrier to dimerization, and demonstrates that the tip-on orientation is more favorable than the face-on one, except at very short distances. Gating of the MscL channel was also studied in Ref.~\cite{Haselwandter13}, by modeling open and closed channels as having different diameters and hydrophobic thicknesses~\cite{Ursell07}. The impact of having different oligomeric states of MscL on these interactions and on gating by tension (see Ref.~\cite{Phillips09}) was studied in Ref.~\cite{Kahraman14}.

The method developed in Ref.~\cite{Haselwandter13} was used in Ref.~\cite{Haselwandter14} to study the effect of membrane-mediated interactions on the self-assembly and architecture of bacterial chemoreceptor lattices. Chemotaxis enables bacteria to perform directed motion in gradients of chemicals. The chemoreceptors that bind to these chemicals are transmembrane proteins that organize into large honeycomb lattices of trimers of dimers at the poles of bacteria~\cite{Briegel12}. In Ref.~\cite{Haselwandter14}, it was shown that membrane-mediated interactions between chemoreceptor trimers of dimers, modeled as inclusions with three-fold symmetry, correctly predict the structure of the arrays observed in experiments. Indeed, at short distances, the face-on relative orientation of the trimers is favored by these anisotropic interactions. In addition, the collective structure of the honeycomb lattice, studied approximately through the pairwise nearest-neighbor interactions, was shown to be more favorable than other types of aggregates at realistic densities of proteins. Gateway states to this lattice were also predicted, and it was shown that membrane-mediated interactions may contribute to the cooperativity of chemotactic signaling.

\subsection{Numerical studies at the microscopic scale}

Continuum models account for the microscopic degrees of freedom (i.e. the positions and conformations of all molecules involved) in a coarse-grained way, via effective terms in the elastic energy and the associated prefactors. However, even in the absence of a mesoscopic deformation due to hydrophobic mismatch, the presence of an inclusion constrains the configurations accessible to the lipid chains that surround it \cite{Marcelja76,Sintes97,Lague00,May00}. Further insight can thus be gained by treating such microscopic degrees of freedom explicitly, in particular those describing the conformation of the lipid chains. Recent advances in numerical simulations have made such approaches possible. Here, we give a brief overview of such studies. Note that numerical studies focusing on larger-scale features were mentioned earlier. 

Refs.~\cite{Lague00,Lague01} used the lateral density-density response function of the alkyl chains, obtained by molecular dynamics simulations of lipid bilayers, to determine the interaction between ``smooth'' (no anchoring) hard cylinders inserted into the bilayer. Three values were considered for the cylinder radius. For the largest one (9~{\AA}, comparable to that of the gramicidin pore, for instance), the long-range interaction is repulsive for all the lipids studied (DMPC, DPPC, POPC and DOPC), with an additional short-range attraction for DMPC. This study does not discuss how the interaction might vary with the concentration of inclusions. Other studies followed suit \cite{Janosi10,Kik10,Yoo13b,Dunton14}. 

A complete description should in principle combine the effects of hydrophobic mismatch and of these changes in chain order \cite{Marcelja99,Bohinc03}. Such a complete model is currently lacking, due to the theoretical difficulties but also due to the dearth of experimental data that could be used to test and validate it. As in the case of lipid tilt (see Section~\ref{sr_iso}), one can wonder how integrating out these underlying degrees of freedom would affect an effective model written in terms of $u$, what effective boundary conditions non-mismatched inclusions would then impose, and whether such a model would be sufficient. 

\subsection{Experimental studies}

It has proven very difficult to directly measure the interactions between membrane inclusions. 

\subsubsection{Electron microscopy}\label{subsec:FFEM}

First among such attempts were freeze-fracture electron microscopy (FFEM) studies
\cite{Lewis83,Chen73,James73,Abney87} that analyzed the spatial distribution of inclusions to determine their radial distribution function $g(r)$. The data was then described using liquid state theories \cite{Pearson83,Pearson84,Braun87} in terms of a hard-core model with an additional interaction, either repulsive or attractive depending on the system.

These pioneering results were not followed by more systematic investigations, probably due to the intrinsic difficulty of the technique. It is also very difficult to check whether the distribution function observed in the sample after freezing still corresponds to that at thermal equilibrium.

\subsubsection{Atomic force microscopy}\label{subsec:AFM}

\begin{figure}
	\centerline{\includegraphics[width=0.7\textwidth,angle=0]{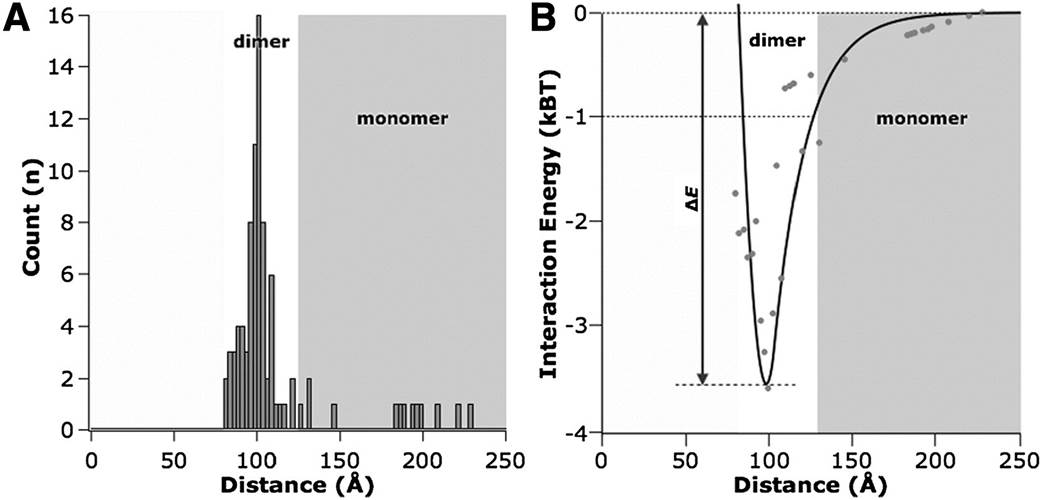}}%
	\caption{\label{fig:Casuso10}Interaction between ATP synthase c-rings. (A) Histogram of the center-to-center distance of c-rings. (B) Membrane-mediated two-protein interaction energy landscape. Reprinted from Figure~2 of reference \cite{Casuso10}.}%
\end{figure}

It has been known for a long time that atomic force microscopy (AFM) can resolve lateral structures down to the nanometer scale \cite{Oesterhelt00}, but data acquisition used to be relatively slow. This changed with the introduction of high-speed AFM \cite{Ando01}, which allows taking ``snapshots'' of the system and determining the radial distribution function. The latter gives access to [OK?] the interaction potential between inclusions, as illustrated by Ref.~\cite{Casuso10} for ATP-synthase c-rings in purple membranes (see Fig.~\ref{fig:Casuso10}). 

\subsubsection{Small-angle scattering}\label{subsec:SAXS}

A promising way of studying membrane-mediated interactions is through small-angle radiation (X-ray or neutrons) scattering from oriented samples, as demonstrated by Refs.~\cite{He95,He96,Yang99}. This non-invasive technique is very well adapted to measurements of membrane-mediated interactions since the wavelength used is of the same order of magnitude as the typical length scales over which one must probe the system (nanometers). One can thus measure the structure factor of the two-dimensional system formed by the inclusions in the membrane and obtain the interaction potential between them.

This strategy was recently used to study alamethicin pores in DMPC membranes \cite{Constantin07}, inorganic particles contained in bilayers of a synthetic surfactant \cite{Constantin08,Constantin10b} and gramicidin pores in several types of membranes \cite{Constantin09}.

\begin{figure}
	\centerline{\includegraphics[width=0.5\textwidth,angle=0]{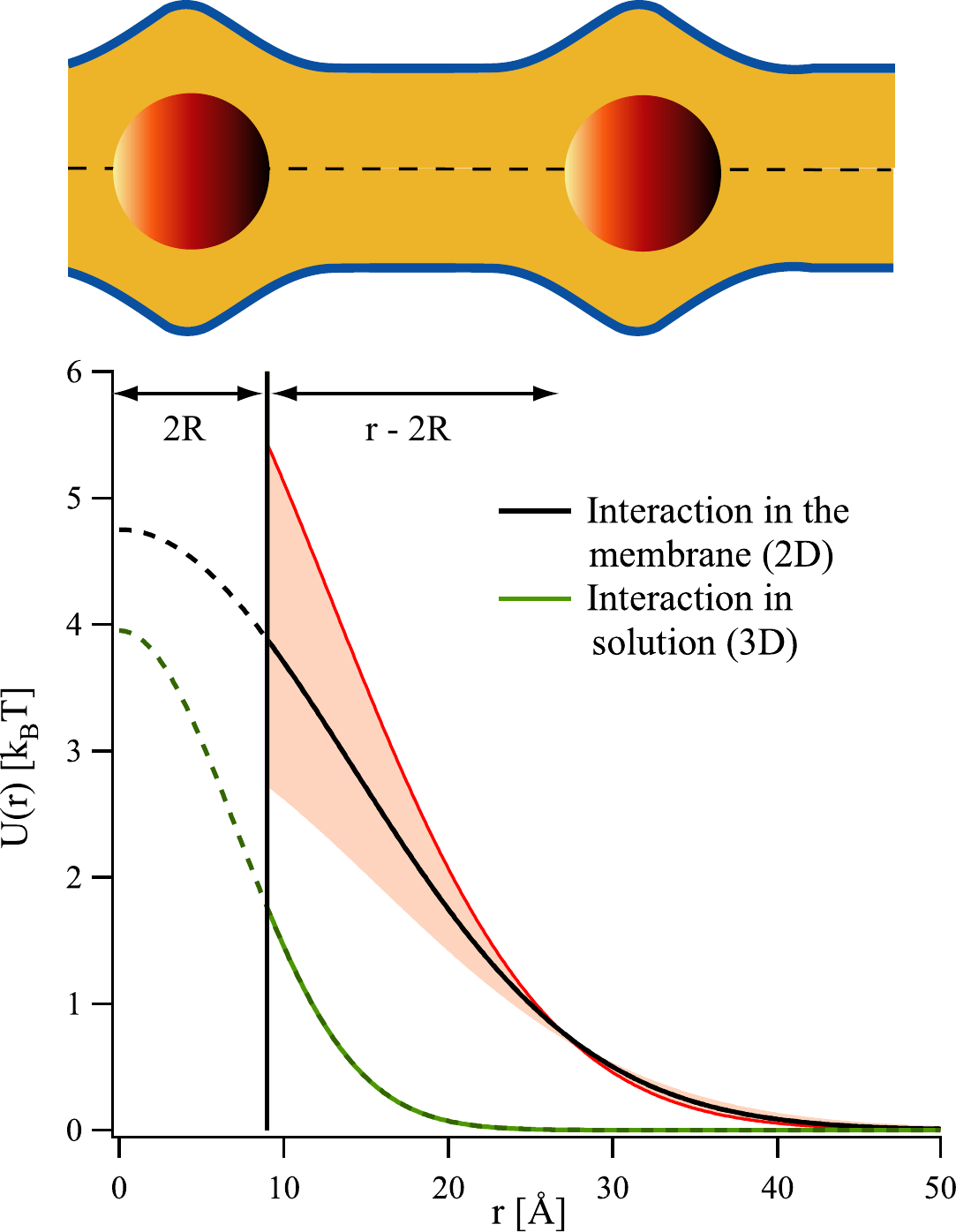}}%
	\caption{\label{fig:BuSn}Interaction potential $U(r)$ of BuSn12 particles within DDAO bilayers. The lower curve is the interaction potential of the particles in ethanol. The solid vertical line marks the hard core interaction with radius 4.5~{\AA}. Reprinted from Figure~3 of reference \cite{Constantin08}.}%
\end{figure}

\section{Conclusion}
\label{sec:Conc}

Membrane-mediated interactions between inclusions constitute a very rich topic. Their study gives insight into the behavior of complex two-dimensional biological membranes. In particular, these interactions may have important impacts on membrane protein aggregation, and on the formation of specific biologically functional assemblies. Interestingly, inclusions can also serve as membrane probes, since membrane-mediated interactions are in part determined by the properties of the host membrane. 

The field of long-range membrane-mediated interactions has been dominated by theory, yielding interesting theoretical developments such as the fluctuation-induced interaction, the general effective field theory and scattering approaches, and the questions currently raised by the dynamics of these interactions. Some experimental and numerical studies have enriched this field, and we hope for further progress allowing for more quantitative comparison with theory.

The study of short-range membrane thickness deformations was motivated by quantitative experiments on gramicidin. Work on these deformations and on the associated membrane-mediated interactions has led to several developments of the theoretical description of membrane elasticity at the nanoscale. Importantly, the small-scale deformations involved are at the limit of the domain of validity of standard coarse-grained continuum theories, making comparison to precise experimental and numerical data even more crucially important. 

An interesting fundamental feature of membrane-mediated interactions is the existence of many-body effects, arising from the interplay of the deformations caused by each of the inclusions. It would thus be particularly interesting to vary the concentration of inclusions in experiments.

\section*{Acknowledgments}
A.-F. B. acknowledges support from the Human Frontier Science Program. D. C. and J.-B. F. acknowledge support by ANR Grant MEMINT (2012-BS04-0023).



\begin{thebibliography}{100}

\bibitem{Singer72}
S.~J. Singer and G.~L. Nicolson.
\newblock The fluid mosaic model of the structure of cell membranes.
\newblock {\em Science}, 175:720--31, 1972.

\bibitem{Sackmann84}
E.~Sackmann.
\newblock Physical basis for trigger processes and membrane structures.
\newblock In D.~Chapman, editor, {\em Biological Membranes}, volume~5, pages
  105--143. Academic Press, London, 1984.

\bibitem{Jensen04}
M.~{\O} Jensen and O.~G. Mouritsen.
\newblock Lipids do influence protein function--the hydrophobic matching
  hypothesis revisited.
\newblock {\em Biochim. Biophys. Acta-Biomembranes}, 1666:205--226, 2004.

\bibitem{Helfrich73}
W.~Helfrich.
\newblock {Elastic properties of lipid bilayers -- Theory and possible
  experiments}.
\newblock {\em {Zeitschrift f\"ur Naturforschung C -- Journal of Biosciences}},
  {28}:693--703, {1973}.

\bibitem{Goulian93etc}
M.~Goulian, R.~Bruinsma, and P.~Pincus.
\newblock {Long-range forces in heterogeneous fluid membranes}.
\newblock {\em EPL}, {22}({2}):145--150, {1993}.
\newblock {Erratum: \textit{Europhysics Letters} 23(2):155, 1993. Comment:
  Ref.~\cite{Fournier97comment}. A factor 2 in the fluctuation-induced force
  was corrected in Ref.~\cite{Golestanian96b}.}

\bibitem{Fournier15}
J.-B. Fournier and P.~Galatola.
\newblock Higher-order power series expansion of the elastic interaction
  between conical membrane inclusions.
\newblock {\em Eur. Phys. J. E}, 38:86, Aug 2015.

\bibitem{Fournier97comment}
J.-B. Fournier and P.~G. Dommersnes.
\newblock {Comment on ``Long-range forces in heterogeneous fluid membranes''}.
\newblock {\em EPL}, {39}({6}):681--682, {1997}.

\bibitem{Golestanian96b}
R.~Golestanian, M.~Goulian, and M.~Kardar.
\newblock {Fluctuation-induced interactions between rods on a membrane}.
\newblock {\em Phys. Rev. E}, {54}({6}):6725--6734, {1996}.

\bibitem{Park96}
J.~M. Park and T.~C. Lubensky.
\newblock {Interactions between membrane inclusions on fluctuating membranes}.
\newblock {\em J. Phys. I}, {6}({9}):1217--1235, {1996}.

\bibitem{Netz97}
R.~R. Netz.
\newblock {Inclusions in fluctuating membranes: exact results}.
\newblock {\em {Journal de Physique I -- France}}, {7}:{833--852}, {1997}.

\bibitem{Kim98}
K.~S. Kim, J.~Neu, and G.~Oster.
\newblock {{C}urvature-mediated interactions between membrane proteins}.
\newblock {\em Biophys. J.}, 75(5):2274--2291, Nov 1998.

\bibitem{Dommersnes99}
P.~G. Dommersnes and J.-B. Fournier.
\newblock {N-body study of anisotropic membrane inclusions: Membrane mediated
  interactions and ordered aggregation}.
\newblock {\em Eur. Phys. J. B}, {12}({1}):9--12, {1999}.

\bibitem{Dommersnes99b}
P.~G. Dommersnes and J.-B. Fournier.
\newblock {Casimir and mean-field interactions between membrane inclusions
  subject to external torques}.
\newblock {\em EPL}, {46}({2}):256--261, {1999}.

\bibitem{Yolcu11}
C.~Yolcu, I.~Z. Rothstein, and M.~Deserno.
\newblock {Effective field theory approach to Casimir interactions on soft
  matter surfaces}.
\newblock {\em EPL}, {96}:{20003}, {2011}.

\bibitem{Yolcu12}
C.~Yolcu and M.~Deserno.
\newblock {Membrane-mediated interactions between rigid inclusions: an
  effective field theory}.
\newblock {\em Phys. Rev. E}, {86}:{031906}, {2012}.

\bibitem{Yolcu12b}
C.~Yolcu, I.~Z. Rothstein, and M.~Deserno.
\newblock {Effective field theory approach to fluctuation-induced forces
  between colloids at an interface}.
\newblock {\em Phys. Rev. E}, {85}:{011140}, {2012}.

\bibitem{Yolcu14}
C.~Yolcu and M.~Deserno.
\newblock {The Effective Field Theory approach towards membrane-mediated
  interactions between particles}.
\newblock {\em {Adv. Colloid Interface Sci.}}, {208}:{89--109}, {2014}.

\bibitem{Golestanian96c}
R.~Golestanian.
\newblock {Reduced persistence length and fluctuation-induced interactions of
  directed semiflexible polymers on fluctuating surfaces}.
\newblock {\em EPL}, {36}:557, {1996}.

\bibitem{Casimir48}
H.~B.~G. Casimir.
\newblock {On the attraction between two perfectly conducting plates}.
\newblock {\em Proc. K. Ned. Akad. Wet.}, {51}({7}):793--796, {1948}.

\bibitem{Milonni}
P.~W. Milonni.
\newblock {\em The Quantum Vacuum -- An Introduction to Quantum
  Electrodynamics}.
\newblock Academic Press, 1994.

\bibitem{Helfrich01}
W.~Helfrich and T.R. Weikl.
\newblock {Two direct methods to calculate fluctuation forces between rigid
  objects embedded in fluid membranes}.
\newblock {\em Eur. Phys. J. E}, {5}:{423--439}, {2001}.

\bibitem{Kardar99}
M.~Kardar and R.~Golestanian.
\newblock {The ``friction'' of vacuum, and other fluctuation-induced forces}.
\newblock {\em Rev. Mod. Phys.}, {71}({4}):1233--1245, {1999}.

\bibitem{Gambassi09}
A.~Gambassi.
\newblock {The Casimir effect: from quantum to critical fluctuations}.
\newblock {\em {JPCS}}, {161}:012037, {2009}.

\bibitem{Fisher78}
M.~E. Fisher and P.~G. de~Gennes.
\newblock {Wall phenomena in a critical binary mixture}.
\newblock {\em C. R. S\'eances Acad. Sci. - Ser. B}, {287}({8}):207--209,
  {1978}.

\bibitem{Hertlein08}
C.~Hertlein, L.~Helden, A.~Gambassi, S.~Dietrich, and C.~Bechinger.
\newblock {Direct measurement of critical Casimir forces}.
\newblock {\em {Nature}}, {451}({7175}):172--175, {2008}.

\bibitem{Machta12}
B.~B. Machta, S.~L. Veatch, and J.~P. Sethna.
\newblock Critical casimir forces in cellular membranes.
\newblock {\em Phys. Rev. Lett.}, 109:138101, Sep 2012.

\bibitem{Golestanian96a}
R.~Golestanian, M.~Goulian, and M.~Kardar.
\newblock {Fluctuation-induced interactions between rods on membranes and
  interfaces}.
\newblock {\em EPL}, {33}({3}):241--245, {1996}.

\bibitem{Stamou00}
D.~Stamou, C.~Duschl, and D.~Johannsmann.
\newblock {Long-range attraction between colloidal spheres at the air-water
  interface: The consequence of an irregular meniscus}.
\newblock {\em Phys. Rev. E}, {62}({4, B}):{5263--5272}, {Oct} {2000}.

\bibitem{Dommersnes02}
P.~G. Dommersnes and J.-B. Fournier.
\newblock {The many-body problem for anisotropic membrane inclusions and the
  self-assembly of ``saddle'' defects into an ``egg carton''}.
\newblock {\em Biophys. J.}, {83}:2898--2905, {2002}.

\bibitem{Axilrod}
B.~M. Axilrod and E.~Teller.
\newblock {Interaction of the van der Waals type between three atoms}.
\newblock {\em J. Chem. Phys.}, {11}({6}):{299--300}, {1943}.

\bibitem{Simunovic13}
M.~Simunovic, A.~Srivastava, and G.~A. Voth.
\newblock {Linear aggregation of proteins on the membrane as a prelude to
  membrane remodeling}.
\newblock {\em Proc. Natl. Acad. Sci. USA}, {110}({51}):{20396--20401}, {Dec}
  {2013}.

\bibitem{Destainville08}
N.~Destainville.
\newblock {Cluster phases of membrane proteins}.
\newblock {\em Phys. Rev. E}, {77}:011905, {1998}.

\bibitem{Weikl01}
T.~R. Weikl.
\newblock {Fluctuation-induced aggregation of rigid membrane inclusions}.
\newblock {\em EPL}, {54}:{547--553}, {2001}.

\bibitem{Rawicz00}
W.~Rawicz, K.~C. Olbrich, T.~McIntosh, D.~Needham, and E.~Evans.
\newblock Effect of chain length and unsaturation on elasticity of lipid
  bilayers.
\newblock {\em Biophys. J.}, 79:328--339, 2000.

\bibitem{Lehle07}
H.~Lehle and M.~Oettel.
\newblock {Importance of boundary conditions for fluctuation-induced forces
  between colloids at interfaces}.
\newblock {\em Phys. Rev. E}, {75}:011602, {2007}.

\bibitem{Noruzifar09}
E.~Noruzifar and M.~Oettel.
\newblock {Anisotropies in thermal Casimir interactions: Ellipsoidal colloids
  trapped at a fluid interface}.
\newblock {\em Phys. Rev. E}, 79(5):051401, 2009.

\bibitem{Lin11}
H.-K. Lin, R.~Zandi, U.~Mohideen, and L.~P Pryadko.
\newblock {Fluctuation-induced forces between inclusions in a fluid membrane
  under tension}.
\newblock {\em Phys. Rev. Lett.}, {107}:228104, {2011}.

\bibitem{Genet03}
C.~Genet, A.~Lambrecht, and S.~Reynaud.
\newblock {Casimir force and the quantum theory of lossy optical cavities}.
\newblock {\em {Phys. Rev. A}}, {67}({4}):{ 043811}, {Apr} {2003}.

\bibitem{Emig08}
T.~Emig, N.~Graham, R.~L. Jaffe, and M.~Kardar.
\newblock {Casimir forces between compact objects: The scalar case}.
\newblock {\em Phys. Rev. D}, {77}({2}):025005, {Jan} {2008}.

\bibitem{Rahi09}
S.~J. Rahi, T.~Emig, N.~Graham, R.~L. Jaffe, and Kardar M.
\newblock {Scattering theory approach to electrodynamic Casimir forces}.
\newblock {\em Phys. Rev. D}, {80}:{085021}, {2009}.

\bibitem{Weikl98}
T.~R. Weikl, M.~M. Kozlov, and W.~Helfrich.
\newblock {Interaction of conical membrane inclusions: Effect of lateral
  tension}.
\newblock {\em Phys. Rev. E}, {57}:{6988--6995}, {1998}.

\bibitem{Evans03}
AR~Evans, MS~Turner, and P~Sens.
\newblock {Interactions between proteins bound to biomembranes}.
\newblock {\em Phys. Rev. E}, {67}:{041907}, {Apr} {2003}.

\bibitem{Simunovic15}
M.~Simunovic and G.~A. Voth.
\newblock {Membrane tension controls the assembly of curvature-generating
  proteins}.
\newblock {\em Nat. Commun.}, {6}:{7219}, {May} {2015}.

\bibitem{Bitbol10}
A.-F. Bitbol, P.~G. Dommersnes, and J.-B. Fournier.
\newblock {Fluctuations of the Casimir-like force between two membrane
  inclusions}.
\newblock {\em Phys. Rev. E}, {81}:050903(R), {2010}.

\bibitem{Capovilla02}
R.~Capovilla and J.~Guven.
\newblock {Stresses in lipid membranes}.
\newblock {\em J. Phys. A}, {35}({30}):6233--6247, {2002}.

\bibitem{Fournier07}
J.-B. Fournier.
\newblock {On the stress and torque tensors in fluid membranes}.
\newblock {\em {Soft Matter}}, {3}({7}):883--888, {2007}.

\bibitem{Barton91}
G.~Barton.
\newblock {On the fluctuations of the Casimir force}.
\newblock {\em J. Phys. A}, {24}({5}):991--1005, {1991}.

\bibitem{Bartolo02}
D.~Bartolo, A.~Ajdari, J.-B. Fournier, and R.~Golestanian.
\newblock {Fluctuations of fluctuation-induced Casimir-like forces}.
\newblock {\em Phys. Rev. Lett.}, {89}({23}):230601, {2002}.

\bibitem{Bitbol11_force}
A.-F. Bitbol and J.-B. Fournier.
\newblock {Forces exerted by a correlated fluid on embedded inclusions}.
\newblock {\em Phys. Rev. E}, {83}:{061107}, {2011}.

\bibitem{Fournier14}
J.-B. Fournier.
\newblock Dynamics of the force exchanged between membrane inclusions.
\newblock {\em Phys. Rev. Lett.}, 112:128101, Mar 2014.
\newblock {Erratum: \textit{Phys. Rev. Lett.}, 114:219901, 2015.}

\bibitem{Seifert93}
U.~Seifert and S.~A. Langer.
\newblock {Viscous modes of fluid bilayer membranes}.
\newblock {\em EPL}, {23}:71--76, {1993}.

\bibitem{Evans94}
E.~Evans and A.~Yeung.
\newblock {Hidden dynamics in rapid changes of bilayer shape}.
\newblock {\em Chem. Phys. Lipids}, {73}:39--56, {1994}.

\bibitem{Fournier15b}
J.-B. Fournier.
\newblock {On the hydrodynamics of bilayer membranes}.
\newblock {\em {Int. J. Nonlinear Mech.}}, {75}:{67--76}, {Oct} {2015}.

\bibitem{Bitbol11_stress}
A.-F. Bitbol, L.~Peliti, and J.-B. Fournier.
\newblock {Membrane stress tensor in the presence of lipid density and
  composition inhomogeneities}.
\newblock {\em Eur. Phys. J. E}, {34}:{53}, {2011}.

\bibitem{Dommersnes98}
P.~G. Dommersnes, J.-B. Fournier, and Galatola P.
\newblock {Long-range elastic forces between membrane inclusions in spherical
  vesicles}.
\newblock {\em EPL}, {42}:233--238, {1998}.

\bibitem{Podgornik95}
R.~Podgornik.
\newblock {Orientational ordering of polymers on a fluctuating flexible
  surface}.
\newblock {\em Phys. Rev. E}, {52}:{5170--5177}, {1995}.

\bibitem{Li92}
H.~Li and M.~Kardar.
\newblock {Fluctuation-induced forces between manifolds immersed in correlated
  fluids}.
\newblock {\em Phys. Rev. A}, {46}({10}):6490--6500, {1992}.

\bibitem{Bitbol11_rods}
A.-F. Bitbol, K.~Sin~Ronia, and J.-B. Fournier.
\newblock {Universal amplitudes of the Casimir-like interactions between four
  types of rods in fluid membranes}.
\newblock {\em EPL}, {96}:{40013}, {2011}.

\bibitem{SinRonia12}
K.~Sin~Ronia and J.-B. Fournier.
\newblock {Universality in the point discretization method for calculating
  Casimir interactions with classical Gaussian fields}.
\newblock {\em {EPL}}, {100}({3}), {Nov} {2012}.

\bibitem{Derjaguin34}
B.~V. Derjaguin.
\newblock {Analysis of friction and adhesion -- IV. The theory of the adhesion
  of small particles}.
\newblock {\em {Kolloid Zeitschrift}}, {69}:155, {1934}.

\bibitem{Weikl03}
T.~R. Weikl.
\newblock {Indirect interactions of membrane-adsorbed cylinders}.
\newblock {\em Eur. Phys. J. E}, {12}:265--273, {2003}.

\bibitem{Muller05}
M.~M. M\"uller, M.~Deserno, and J.~Guven.
\newblock Interface-mediated interactions between particles: A geometrical
  approach.
\newblock {\em Phys. Rev. E}, {72}:061407, {2005}.

\bibitem{Muller05b}
M.~M. M\"uller, M.~Deserno, and J.~Guven.
\newblock Geometry of surface-mediated interactions.
\newblock {\em EPL}, {69}:482--488, {2005}.

\bibitem{Muller07}
M.~M. M\"uller and M.~Deserno.
\newblock Balancing torques in membrane-mediated interactions: Exact results
  and numerical illustrations.
\newblock {\em Phys. Rev. E}, {76}:011921, {2007}.

\bibitem{Mkrtchyan10}
Sergey Mkrtchyan, Christopher Ing, and Jeff Z.~Y. Chen.
\newblock Adhesion of cylindrical colloids to the surface of a membrane.
\newblock {\em Phys. Rev. E}, 81:011904, Jan 2010.

\bibitem{Gosselin11}
P.~Gosselin, H.~Mohrbach, and M.~M. M\"uller.
\newblock {Interface-mediated interactions: Entropic forces of curved
  membranes}.
\newblock {\em Phys. Rev. E}, {83}:051921, {2011}.

\bibitem{Reynwar07}
B.~J. Reynwar, G.~Illya, V.~A. Harmandaris, M.~M. Mueller, K.~Kremer, and
  M.~Deserno.
\newblock {Aggregation and vesiculation of membrane proteins by
  curvature-mediated interactions}.
\newblock {\em {Nature}}, {447}:461--464, {2007}.

\bibitem{Yuan11}
H.~Yuan, C.~Huang, and S.~Zhang.
\newblock {Membrane-Mediated Inter-Domain Interactions}.
\newblock {\em BioNanoSci.}, {1}:97--102, {Jun} {2011}.

\bibitem{Reynwar11}
B.~J. Reynwar and M.~Deserno.
\newblock {Membrane-mediated interactions between circular particles in the
  strongly curved regime}.
\newblock {\em Soft Matter}, {7}:8567, {2011}.

\bibitem{Schweitzer15}
Y.~Schweitzer and M.~M. Kozlov.
\newblock {Membrane-Mediated Interaction between Strongly Anisotropic Protein
  Scaffolds}.
\newblock {\em {PLOS Comp. Biol.}}, {11}({2}), {Feb} {2015}.

\bibitem{Bahrami12}
A.~H. Bahrami, R.~Lipowsky, and T.~R. Weikl.
\newblock {{T}ubulation and aggregation of spherical nanoparticles adsorbed on
  vesicles}.
\newblock {\em Phys. Rev. Lett.}, 109(18):188102, Nov 2012.

\bibitem{Saric13}
A.~\v{S}ari\'{c} and A.~Cacciuto.
\newblock {Self-assembly of nanoparticles adsorbed on fluid and elastic
  membranes}.
\newblock {\em {Soft Matter}}, {9}:6677--6695, {2013}.

\bibitem{Ramakrishnan13}
N.~Ramakrishnan, P.~B. Sunil~Kumar, and J.~H. Ipsen.
\newblock {{M}embrane-mediated aggregation of curvature-inducing nematogens and
  membrane tubulation}.
\newblock {\em Biophys. J.}, 104(5):1018--1028, Mar 2013.

\bibitem{Roemer07}
Winfried R\"{o}mer, Ludwig Berland, Val\'{e}rie Chambon, Katharina Gaus,
  Barbara Windschiegl, Dani\`{e}le Tenza, Mohamed R.~E. Aly, Vincent Fraisier,
  Jean-Claude Florent, David Perrais, Christophe Lamaze, Gra\c{c}ça Raposo,
  Claudia Steinem, Pierre Sens, Patricia Bassereau, and Ludger Johannes.
\newblock Shiga toxin induces tubular membrane invaginations for its uptake
  into cells.
\newblock {\em Nature}, 450:670--675, 2007.

\bibitem{Reynawar07}
Benedict~J. Reynwar, Gregoria Illya, Vagelis~A. Harmandaris, Martin M.Mueller,
  Kurt Kremer, and Markus Deserno.
\newblock Aggregation and vesiculation of membrane proteins by
  curvature-mediated interactions.
\newblock {\em Nature}, 447:461--464, 2007.

\bibitem{McMahon05}
H.~T. McMahon and J.~L. Gallop.
\newblock {Membrane curvature and mechanisms of dynamic cell membrane
  remodelling}.
\newblock {\em {Nature}}, {438}:590--596, {2005}.

\bibitem{Angelova86}
M.~I. Angelova and D.~S. Dimitrov.
\newblock {Liposome electroformation}.
\newblock {\em Faraday Discuss. Chem. Soc.}, {81}:{303--311}, {1986}.

\bibitem{Semrau09}
S.~Semrau, T.~Idema, T.~Schmidt, and C.~Storm.
\newblock {Membrane-mediated interactions measured using membrane domains}.
\newblock {\em Biophys. J.}, {96}:4906–4915, {2009}.

\bibitem{Koltover99}
I.~Koltover, J.~O. R\"adler, and C.~R. Safinya.
\newblock {Membrane mediated attraction and ordered aggregation of colloidal
  particles bound to giant phospholipid vesicles}.
\newblock {\em Phys. Rev. Lett.}, {82}:{1991--1994}, {1999}.

\bibitem{Yue12}
T.~Yue and X.~Zhang.
\newblock {Cooperative Effect in Receptor-Mediated Endocytosis of Multiple
  Nanoparticles}.
\newblock {\em {ACS Nano}}, {6}({4}):{3196--3205}, {Apr} {2012}.

\bibitem{Saric12}
A.~Saric and A.~Cacciuto.
\newblock {Fluid Membranes Can Drive Linear Aggregation of Adsorbed Spherical
  Nanoparticles}.
\newblock {\em Phys. Rev. Lett.}, {108}({11}):118101, {Mar} {2012}.

\bibitem{Rozovsky05}
Sharon Rozovsky, Yoshihisa Kaizuka, and Jay~T. Groves.
\newblock Formation and spatio-temporal evolution of periodic structures in
  lipid bilayers.
\newblock {\em Journal of the American Chemical Society}, 127(1):36--37, 2005.

\bibitem{Ursell09}
Tristan~S. Ursell, William~S. Klug, and Rob Phillips.
\newblock Morphology and interaction between lipid domains.
\newblock {\em Proceedings of the National Academy of Sciences},
  106(32):13301--13306, 2009.

\bibitem{Heberle13}
Frederick~A. Heberle, Robin~S. Petruzielo, Jianjun Pan, Paul Drazba, Norbert
  Ku{\v c}erka, Robert~F. Standaert, Gerald~W. Feigenson, and John Katsaras.
\newblock Bilayer {Thickness} {Mismatch} {Controls} {Domain} {Size} in {Model}
  {Membranes}.
\newblock {\em Journal of the American Chemical Society}, 135(18):6853--6859,
  2013.

\bibitem{Jost73}
P.~C. Jost, O.~H. Griffiths, R.~A. Capaldi, and G.~Vanderkooi.
\newblock {Evidence for boundary lipid in membranes}.
\newblock {\em Proc. Natl. Acad. Sci. USA}, {70}:480--484, {1973}.

\bibitem{Marcelja76}
S.~Mar{\v c}elja.
\newblock {Lipid-mediated protein interaction in membranes}.
\newblock {\em Biochim. Biophys. Acta: Biomembr.}, {455}:{1--7}, {1976}.

\bibitem{Schroder77}
H.~Schr\"oder.
\newblock {Aggregation of proteins in membranes. An example of fluctuation
  induced interactions in liquid crystals}.
\newblock {\em J. Chem. Phys.}, {67}:{1617}, {1977}.

\bibitem{Owicki78}
J.~C. Owicki, M.~W. Springgate, and H.~McConnell.
\newblock {Theoretical study of protein-lipid interactions in bilayer
  membranes}.
\newblock {\em Proc. Natl. Acad. Sci. USA}, {75}:1616--1619, {1978}.

\bibitem{Killian98}
J.~A. Killian.
\newblock {Hydrophobic mismatch between proteins and lipids in membranes}.
\newblock {\em Biochim. Biophys. Acta: Biomembr.}, {1376}:401--416, {1998}.

\bibitem{Mouritsen84}
O.~G. Mouritsen and M.~Bloom.
\newblock {Mattress model of lipid-protein interactions in membranes}.
\newblock {\em Biophys. J.}, {46}:141--153, {1984}.

\bibitem{Kelkar07}
D.~A. Kelkar and A.~Chattopadhyay.
\newblock The gramicidin ion channel: A model membrane protein.
\newblock {\em Biochim. Biophys. Acta: Biomembr.}, 1768:2011--2025, 2007.

\bibitem{Kolb77}
H.~A. Kolb and E.~Bamberg.
\newblock Influence of membrane thickness and ion concentration on the
  properties of the gramicidin {A} channel: autocorrelation, spectral power
  density, relaxation and single-channel studies.
\newblock {\em Biochim. Biophys. Acta: Biomembr.}, 464:127--141, 1977.

\bibitem{Elliott83}
J.~R. Elliott, D.~Needham, J.~P. Dilger, and D.~A. Haydon.
\newblock The effects of bilayer thickness and tension on gramicidin
  single-channel lifetime.
\newblock {\em Biochim. Biophys. Acta: Biomembr.}, 735:95--103, 1983.

\bibitem{Goulian98}
M.~Goulian, O.~N. Mesquita, D.~K. Fygenson, C.~Nielsen, O.~S. Andersen, and
  A.~Libchaber.
\newblock {Gramicidin channel kinetics under tension}.
\newblock {\em Biophys. J.}, {74}:{328--337}, {1998}.

\bibitem{Huang86}
H.~W. Huang.
\newblock {Deformation free energy of bilayer membrane and its effect on
  gramicidin channel lifetime}.
\newblock {\em Biophys. J.}, {50}:{1061--1070}, {1986}.

\bibitem{dG_LC}
P.~G. de~Gennes.
\newblock {\em {The Physics of Liquid Crystals}}.
\newblock {Clarendon Press, Oxford}, {1974}.

\bibitem{Dan93}
N.~Dan, P.~Pincus, and S.~A. Safran.
\newblock {Membrane-induced interactions between inclusions}.
\newblock {\em {Langmuir}}, {9}:{2768--2771}, {1993}.

\bibitem{ArandaEspinoza96}
H.~Aranda-Espinoza, A.~Berman, N.~Dan, P.~Pincus, and S.~Safran.
\newblock {Interaction between inclusions embedded in membranes}.
\newblock {\em Biophys. J.}, {71}:648, {1996}.

\bibitem{Brannigan06}
G.~Brannigan and F.~L.~H. Brown.
\newblock {A consistent model for thermal fluctuations and protein-induced
  deformations in lipid bilayers}.
\newblock {\em Biophys. J.}, {90}({5}):{1501--1520}, {2006}.

\bibitem{Brannigan07}
G.~Brannigan and F.~L.~H. Brown.
\newblock Contributions of gaussian curvature and nonconstant lipid volume to
  protein deformation of lipid bilayers.
\newblock {\em Biophys. J.}, {92}({3}):{864--876}, {2007}.

\bibitem{Watson11}
M.~C. Watson, E.~S. Penev, P.~M. Welch, and F.~L.~H. Brown.
\newblock {Thermal fluctuations in shape, thickness, and molecular orientation
  in lipid bilayers}.
\newblock {\em J. Chem. Phys.}, {135}:{244701}, {2011}.

\bibitem{Bitbol12_grami}
A.-F. Bitbol, D.~Constantin, and J.-B. Fournier.
\newblock {Bilayer elasticity at the nanoscale: the need for new terms}.
\newblock {\em {PLoS ONE}}, 7(11):e48306, 2012.

\bibitem{West09}
B.~West, F.~L.~H. Brown, and F.~Schmid.
\newblock Membrane-protein interactions in a generic coarse-grained model for
  lipid bilayers.
\newblock {\em Biophys. J.}, {96}({1}):{101--115}, {2009}.

\bibitem{Owicki79}
J.~C. Owicki and H.~McConnell.
\newblock {Theory of protein-lipid and protein-protein interactions in bilayer
  membranes}.
\newblock {\em Proc. Natl. Acad. Sci. USA}, {76}:4750--4754, {1979}.

\bibitem{Dan94}
N.~Dan, A.~Berman, P.~Pincus, and S.~A. Safran.
\newblock {Membrane-induced interactions between inclusions}.
\newblock {\em {J. Phys. II France}}, {4}:{1713--1725}, {1994}.

\bibitem{Harroun99_theo}
T.~A. Harroun, W.~T. Heller, T.~M. Weiss, L.~Yang, and H.~W. Huang.
\newblock {{T}heoretical analysis of hydrophobic matching and membrane-mediated
  interactions in lipid bilayers containing gramicidin}.
\newblock {\em Biophys. J.}, 76(6):3176--3185, Jun 1999.

\bibitem{Neder11}
J.~Neder, B.~West, P.~Nielaba, and F.~Schmid.
\newblock {Membrane-mediated Protein-protein Interaction: A Monte Carlo Study}.
\newblock {\em {Curr. Nanosci.}}, {7}({5}):{656--666}, {Oct} {2011}.

\bibitem{Bories_tbp}
F.~Bories, D.~Constantin, J.-B. Fournier, and P.~Galatola.
\newblock {{To be published.}}

\bibitem{ChaikinLubensky}
P.~M. Chaikin and T.~C. Lubensky.
\newblock {\em Principles of condensed matter physics}.
\newblock Cambridge University Press, 1995.

\bibitem{Fournier98}
J.-B. Fournier.
\newblock {Coupling between membrane tilt-difference and dilation: A new
  ``ripple" instability and multiple crystalline inclusions phases}.
\newblock {\em {EPL}}, 43(6):725--730, 1998.

\bibitem{May99}
S.~May and A.~Ben-Shaul.
\newblock {Molecular theory of lipid-protein interaction and the
  L$_\alpha$-H$_{||}$ transition}.
\newblock {\em Biophys. J.}, {76}:{751--767}, {1999}.

\bibitem{May00}
S.~May and A.~{Ben-Shaul}.
\newblock A molecular model for lipid-mediated interaction between proteins in
  membranes.
\newblock {\em Phys. Chem. Chem. Phys.}, 2:4494--4502, 2000.

\bibitem{May02}
S~May.
\newblock {Membrane perturbations induced by integral proteins: Role of
  conformational restrictions of the lipid chains}.
\newblock {\em Langmuir}, {18}:{6356--6364}, {Aug} {2002}.

\bibitem{Bohinc03}
Klemen Bohinc, Veronika Kralj-Igli{\v{c}}, and Silvio May.
\newblock Interaction between two cylindrical inclusions in a symmetric lipid
  bilayer.
\newblock {\em J. Chem. Phys.}, 119:7435--7444, 2003.

\bibitem{Kozlovsky04}
Y.~Kozlovsky, J.~Zimmerberg, and M.~M. Kozlov.
\newblock {Orientation and interaction of oblique cylindrical inclusions
  embedded in a lipid monolayer: A theoretical model for viral fusion
  peptides}.
\newblock {\em Biophys. J.}, {87}:{999--1012}, {Aug} {2004}.

\bibitem{Haselwandter13}
C.~A. Haselwandter and R.~Phillips.
\newblock Directional interactions and cooperativity between mechanosensitive
  membrane proteins.
\newblock {\em EPL}, 101(6):68002, 2013.

\bibitem{Ursell07}
T.~Ursell, K.~C. Huang, E.~Peterson, and R.~Phillips.
\newblock {Cooperative gating and spatial organization of membrane proteins
  through elastic interactions}.
\newblock {\em {PLoS Comput. Biol.}}, {3}({5}):{803--812}, {May} {2007}.

\bibitem{Phillips09}
R.~Phillips, T.~Ursell, P.~Wiggins, and P.~Sens.
\newblock {Emerging roles for lipids in shaping membrane-protein function}.
\newblock {\em {Nature}}, {459}({7245}):{379--385}, {May} {2009}.

\bibitem{Kahraman14}
Osman Kahraman, William~S. Klug, and Christoph~A. Haselwandter.
\newblock {Signatures of protein structure in the cooperative gating of
  mechanosensitive ion channels}.
\newblock {\em {EPL}}, {107}({4}), {Aug} {2014}.

\bibitem{Haselwandter14}
C.~A. Haselwandter and N.~S. Wingreen.
\newblock {{T}he role of membrane-mediated interactions in the assembly and
  architecture of chemoreceptor lattices}.
\newblock {\em PLoS Comput. Biol.}, 10(12):e1003932, Dec 2014.

\bibitem{Briegel12}
A.~Briegel, X.~Li, A.~M. Bilwes, K.~T. Hughes, G.~J. Jensen, and B.~R. Crane.
\newblock {Bacterial chemoreceptor arrays are hexagonally packed trimers of
  receptor dimers networked by rings of kinase and coupling proteins}.
\newblock {\em Proc. Natl. Acad. Sci. USA}, {109}({10}):{3766--3771}, {Mar}
  {2012}.

\bibitem{Sintes97}
T.~Sintes and A.~Baumg\"{a}rtner.
\newblock Protein attraction in membranes induced by lipid fluctuations.
\newblock {\em Biophys. J.}, 73:2251--2259, 1997.

\bibitem{Lague00}
Patrick Lag{\"{u}}e, Martin~J. Zuckermann, and Beno\^{i}t Roux.
\newblock Lipid-mediated interactions between intrinsic membrane proteins: A
  theoretical study based on integral equations.
\newblock {\em Biophys. J.}, 79:2867--2879, 2000.

\bibitem{Lague01}
Patrick Lag{\"{u}}e, Martin~J. Zuckermann, and Beno\^{i}t Roux.
\newblock Lipid-mediated interactions between intrinsic membrane proteins:
  Dependence on protein size and lipid composition.
\newblock {\em Biophys. J.}, 81:276--284, 2001.

\bibitem{Janosi10}
Lorant Janosi, Anupam Prakash, and Manolis Doxastakis.
\newblock Lipid-{Modulated} {Sequence}-{Specific} {Association} of
  {Glycophorin} {A} in {Membranes}.
\newblock {\em Biophysical Journal}, 99(1):284--292, 2010.

\bibitem{Kik10}
Richard~A. Kik, Frans A.~M. Leermakers, and J.~Mieke Kleijn.
\newblock Molecular modeling of proteinlike inclusions in lipid bilayers:
  {Lipid}-mediated interactions.
\newblock {\em Physical Review E}, 81(2), 2010.

\bibitem{Yoo13b}
Jejoong Yoo and Qiang Cui.
\newblock Membrane-mediated protein-protein interactions and connection to
  elastic models: A coarse-grained simulation analysis of gramicidin {A}
  association.
\newblock {\em Biophysical journal}, 104:128--138, January 2013.

\bibitem{Dunton14}
Thomas~A. Dunton, Joseph~E. Goose, David~J. Gavaghan, Mark S.~P. Sansom, and
  James~M. Osborne.
\newblock The {Free} {Energy} {Landscape} of {Dimerization} of a {Membrane}
  {Protein}, {NanC}.
\newblock {\em PLoS Computational Biology}, 10(1):e1003417, 2014.

\bibitem{Marcelja99}
S.~{Mar\v{c}elja}.
\newblock Toward a realistic theory of the interaction of membrane inclusions.
\newblock {\em Biophys. J.}, 76:593--594, 1999.

\bibitem{Lewis83}
B.~A. Lewis and D.~M. Engelman.
\newblock Bacteriorhodopsin remains dispersed in fluid phospholipid bilayers
  over a wide range of bilayer thicknesses.
\newblock {\em J. Mol. Biol.}, 166:203--210, 1983.

\bibitem{Chen73}
Y.~S. Chen and W.~L. Hubbell.
\newblock Temperature- and light-dependent structural changes in
  rhodopsin-lipid membranes.
\newblock {\em Exp. Eye Res}, 17:517--532, 1973.

\bibitem{James73}
R.~James and D.~Branton.
\newblock Lipid- and temperature-dependent structural changes in
  \textit{{Acholeplasma} laidlawii} cell membranes.
\newblock {\em Biochim. Biophys. Acta}, 323:378--390, 1973.

\bibitem{Abney87}
James~R. Abney, Jochen Braun, and John~C. Owicki.
\newblock Lateral interactions among membrane proteins: Implications for the
  organization of gap junctions.
\newblock {\em Biophys. J.}, 52:441--454, 1987.

\bibitem{Pearson83}
L.~T. Pearson, B.~A. Lewis, D.~M. Engelman, and S.~I. Chan.
\newblock Pair distribution functions of bacteriorhodopsin and rhodopsin in
  model bilayers.
\newblock {\em Biophys. J.}, 43:167--174, 1983.

\bibitem{Pearson84}
L.~T. Pearson, J.~Edelman, and S.~I. Chan.
\newblock Statistical mechanics of lipid membranes, protein correlation
  functions and lipid ordering.
\newblock {\em Biophys. J.}, 45:863--871, 1984.

\bibitem{Braun87}
Jochen Braun, James~R. Abney, and John~C. Owicki.
\newblock Lateral interactions among membrane proteins: Valid estimates based
  on freeze-fracture electron microscopy.
\newblock {\em Biophys. J.}, 52:427--439, 1987.

\bibitem{Oesterhelt00}
F.~Oesterhelt, D.~Oesterhelt, M.~Pfeiffer, A.~Engel, H.~E. Gaub, and D.~J.
  Muller.
\newblock Unfolding pathways of individual bacteriorhodopsins.
\newblock {\em Science}, 288:143--146, 2000.

\bibitem{Ando01}
Toshio Ando, Noriyuki Kodera, Eisuke Takai, Daisuke Maruyama, Kiwamu Saito, and
  Akitoshi Toda.
\newblock A high-speed atomic force microscope for studying biological
  macromolecules.
\newblock {\em Proceedings of the National Academy of Sciences},
  98(22):12468--12472, 2001.

\bibitem{Casuso10}
Ignacio Casuso, Pierre Sens, Felix Rico, and Simon Scheuring.
\newblock Experimental evidence for membrane-mediated protein-protein
  interaction.
\newblock {\em Biophysical Journal}, 99(7):L47--L49, 2010.

\bibitem{He95}
K.~He, S.~J. Ludtke, H.~W. Huang, and D.~L. Worcester.
\newblock Antimicrobial peptide pores in membranes detected by neutron in-plane
  scattering.
\newblock {\em Biochem.}, 34:15614--15618, 1995.

\bibitem{He96}
K.~He, S.~J. Ludtke, D.~L. Worcester, and H.~W. Huang.
\newblock Neutron scattering in the plane of the membranes: Structure of
  alamethicin pores.
\newblock {\em Biophys. J.}, 70:2659--2666, 1996.

\bibitem{Yang99}
L.~Yang, T.M. Weiss, T.A. Harroun, W.T. Heller, and H.W. Huang.
\newblock Supramolecular structures of peptide assemblies in membranes by
  neutron off-plane scattering: Method of analysis.
\newblock {\em Biophys. J.}, 77:2648--2656, 1999.

\bibitem{Constantin07}
D.~Constantin, G.~Brotons, A.~Jarre, C.~Li, and T.~Salditt.
\newblock Interaction of alamethicin pores in {DMPC} bilayers.
\newblock {\em Biophys. J.}, 92:3978--3987, 2007.

\bibitem{Constantin08}
Doru Constantin, Brigitte Pansu, Marianne Imp\'{e}ror, Patrick Davidson, and
  Fran\c{c}ois Ribot.
\newblock Repulsion between inorganic particles inserted within surfactant
  bilayers.
\newblock {\em Physical Review Letters}, 101:098101, 2008.

\bibitem{Constantin10b}
Doru Constantin.
\newblock The interaction of hybrid nanoparticles inserted within surfactant
  bilayers.
\newblock {\em The Journal of Chemical Physics}, 133:144901, 2010.

\bibitem{Constantin09}
D.~Constantin.
\newblock {Membrane-mediated repulsion between gramicidin pores}.
\newblock {\em Biochim. Biophys. Acta: Biomembr.}, {1788}:1782--1789, {2009}.

\end{thebibliography}
\end{document}